\documentclass{aa} 
\usepackage{graphicx}
\usepackage{xcolor}
\usepackage{float}
\usepackage{chngcntr}
\usepackage{textcomp}
\usepackage{siunitx}
\usepackage{wrapfig}

\usepackage[colorlinks]{hyperref}  
\hypersetup{citecolor=blue} 

\usepackage{txfonts}

\begin{document}

   \title{Mapping Wolf-Rayet winds with spectro-interferometry}

   \subtitle{I. Detection of multiple co-rotating interaction regions in WR~6}

    \author{K.\ Deshmukh\inst{\ref{inst:kul},\ref{inst:lgi},}\thanks{Corresponding author; \texttt{astro.kunal.deshmukh@gmail.com}} \and 
           A.-N. Chen\'e \inst{\ref{inst:noir}}
           \and
           A. J. Frost \inst{\ref{inst:eso-c}}
           \and 
           T. De Prins \inst{\ref{inst:kul},\ref{inst:mac},\ref{inst:astrc}}
           \and
           L. Fabiani \inst{\ref{inst:mont}}
           \and
           H. Sana \inst{\ref{inst:kul},\ref{inst:lgi}}
           \and 
           R. R. Lefever\inst{\ref{inst:ari}}
           \and \\
           A. Mérand \inst{\ref{inst:eso-g}}
           \and 
           N. D. Richardson \inst{\ref{inst:erau}}
           \and
           A. A. C. Sander \inst{\ref{inst:cau},\ref{inst:ari}}
           \and
           T. Shenar \inst{\ref{inst:tau}}
           }

   \institute{
   {Institute of Astronomy, KU Leuven, Celestijnenlaan 200D, 3001 Leuven, Belgium \label{inst:kul}}\\ 
   \email{astro.kunal.deshmukh@gmail.com}
   \and 
    {Leuven Gravity Institute, KU Leuven, Celestijnenlaan 200D, box 2415, 3001 Leuven, Belgium \label{inst:lgi}}
    \and 
    {NSF NOIRLab, 670 N. A`ohoku Place, Hilo, HI 96720, USA \label{inst:noir}}
    \and 
    {European Southern Observatory, Alonso de Cordova 3107, Vitacura, Santiago, Chile \label{inst:eso-c}}
    \and 
    {School of Mathematical \& Physical Sciences, Macquarie University, Macquarie Park, NSW, Australia \label{inst:mac}}
    \and 
    {Astrophysics and Space Technologies Research Centre, Macquarie University, Macquarie Park, NSW, Australia \label{inst:astrc}}
    \and 
    {D\'epartement de Physique, Universit\'e de Montr\'eal, C.P. 6128, Succ. C-V, Montr\'eal, QC H3C 3J7, Canada \label{inst:mont}}
    \and 
    {Zentrum für Astronomie der Universität Heidelberg, Astronomisches Rechen-Institut, Mönchhofstr. 12-14, 69120 Heidelberg, Germany \label{inst:ari}}
    \and 
    {European Southern Observatory, Karl-Schwarzschild-Straße 2, 85748 Garching, Germany \label{inst:eso-g}}
    \and 
    {Department of Physics and Astronomy, Embry-Riddle Aeronautical University, 3700 Willow Creek Rd, Prescott, AZ 86301, USA\label{inst:erau}}
    \and 
    {Institut f{\"u}r Theoretische Physik und Astrophysik, Christian-Albrechts-Universit{\"a}t zu Kiel, Leibnizstr.\ 15, 24118 Kiel, Germany\label{inst:cau}}
    \and 
    {The School of Physics and Astronomy, Tel Aviv University, Tel Aviv 6997801, Israel \label{inst:tau}}
    }

   \date{Received; Accepted}

  \abstract
   {WR~6 is the prototype for periodic variability due to large-scale wind structures in Wolf-Rayet (WR) stars, showing a consistent periodicity of 3.76~d in photometry, spectroscopy, and polarimetry. However, the origin of this variability is debated, attributed either to corotating interaction regions (CIRs) or a low-mass companion.}
   {We use time-resolved long-baseline interferometry to investigate the origin of periodic variability in WR~6. Inspired by recent interferometric observations that spatially resolved the stellar winds of WR stars, we aim to detect and characterize structures in the stellar winds of WR~6 and discriminate between the CIR and low-mass companion scenarios.}
   {Using the extended configuration of the Auxiliary Telescopes at VLTI, we obtained six GRAVITY epochs over six months to search for the presence and variability of spatially resolved wind-structures in WR~6. We also obtained complementary spectroscopic observations during one of the epochs to monitor contemporaneous spectral line variability.}
   {For all six epochs of GRAVITY observations: (i) WR~6 is consistent with a single star and has no spatially resolved companion detected above our sensitivity limit of $M\gtrsim6\,M_\odot$; (ii) The He\,{\sc i} line-emitting region is spatially resolved, while the He\,{\sc ii} lines remain unresolved, consistent with atmosphere models for WR~6; (iii) Around 2.19 $\mathrm{\mu}$m, one or more discrete line-emitting features offset from the center of the star are detected. The number, positions, and wavelengths of these features vary from epoch to epoch. Spectroscopic monitoring of WR~6 around the time of the fifth epoch reveals spectral line residuals that broadly correspond to the two features seen in the GRAVITY data of the fifth epoch. Finally, the first three epochs, which were observed on three consecutive nights, show significant variability in these features but cannot be directly linked across epochs.}
   {We detect multiple large-scale structures in the wind of WR~6 with VLTI/GRAVITY. We favor CIRs as their likely origin owing to temporal and spatial variability that cannot be reconciled with a low-mass companion. Building upon the first VLTI detection of CIRs, we encourage future campaigns with higher cadence to constrain the time-evolution and geometry of CIRs.}


   \titlerunning{Detecting CIRs in WR~6 with VLTI/GRAVITY}
   
   \maketitle
%

\section{Introduction}

Massive stars profoundly influence the structure, chemistry, and dynamics of their host galaxies by injecting large amounts of energy, momentum, and chemically enriched material into the interstellar medium \citep{2005Kennicutt}. Among them, classical Wolf-Rayet (WR) stars represent an advanced evolutionary stage of stars initially more massive than $\gtrsim$\,20\,M$_\odot$ and are considered direct progenitors of stellar-mass black holes \citep{2019Woosley,2012Langer}. Having shed their hydrogen-rich envelopes, either through powerful radiatively driven winds \citep{1976Conti,2014Smith} or via binary interactions \citep{1967Paczynski,2020Shenar}, classical WR stars expose nuclear-processed material at their surfaces and launch dense, optically thick winds. These winds are far from homogeneous: they exhibit structure over a wide range of spatial and temporal scales, from stochastic clumping to coherent, large-scale perturbations \citep{2009StLouis,2020Chene}.

Among the most striking large-scale wind structures are Co-rotating Interaction Regions (CIRs). First identified in the solar wind and later adapted to hot-star winds \citep{1984Mullan,1996Cranmer}, CIRs arise from persistent surface inhomogeneities whose rotational modulation imprints spiral-shaped density and velocity perturbations into the outflow. In O-type stars, CIRs are observed through Discrete/Narrow Absorption Components (DACs/NACs) in ultraviolet resonance lines \citep{1989Howarth}. Their geometry can be described as Archimedean spirals whose curvature depends on the ratio between wind terminal velocity and stellar rotation rate.

Although small-scale clumping is ubiquitous in WR winds \citep[][and references therein]{2020Chene}, CIRs represent an additional, organized perturbation rooted at or near the onset of the stellar wind. Observational evidence for CIRs in WR stars has been reported in a limited but growing number of cases. In particular, WR~6 (EZ\,CMa) has long been regarded as the prototypical example of a WR star exhibiting persistent, large-scale variability compatible with CIR modulation. Periodic spectroscopic and photometric variations with a timescale of 3.76\,days were first reported by \citet{1980Firmani}. \citet{1997Morel} proposed that these variations could be explained by CIRs embedded in the wind, an interpretation later supported for other WR stars such as WR~1 \citep{2010Chene,2013StLouis}, WR~110 \citep{2011Chene}, and WR~134 \citep{1999bMorel,2016Aldoretta}. More recently, spectropolarimetric observations have revealed loops in the $Q$–$U$ plane \citep{2026Fabiani} consistent with occultation effects produced by rotating CIR structures in WR~6, in agreement with the models of \citet{2023Ignace}; a similar behavior has also been detected in WR~1 \citep{RNAAS2025}.

All WR stars exhibiting WR~6-like photometric and spectroscopic variability belong to the WN[4–6]b subtype, raising the possibility that CIRs are preferentially found in this class. If so, these objects may not be exceptional, but instead representatives of a subset of WR stars whose winds are shaped by stable, large-scale surface perturbations. Understanding the origin, geometry, and physical properties of CIRs in WR winds therefore allows us to constrain the dynamics of optically thick, radiatively driven outflows and the nature of surface inhomogeneities in evolved massive stars.

As an alternative to the CIR scenario, binarity has repeatedly been invoked to explain the variability of WR~6. The original discovery of the 3.76-day periodicity led \citet{1980Firmani} to suggest a WR + neutron star(NS) system. However, the absence of the expected hard X-ray emission from accretion, even if justifiable by a NS orbiting inside the optically thick WR wind \citep[see ][]{2018Toala}, weakened this interpretation \citep{2002Skinner}. More recently, \citet{2019SchmutzKoenigsbergerWR6} and \citet{2020Koenigsberger} proposed a close, low-mass, non-degenerate companion whose interaction with the WR wind would generate a localized high-density structure, mimicking CIR-like signatures. In this configuration, the companion would orbit at $a = 0.13$~AU, and the observed variability would arise from wind–companion interaction rather than intrinsic surface perturbations. However, the presence of such a companion could not be established using high-resolution spectroscopic observations without invoking an extremely high apsidal motion rate \citep{2024Barclay}.

Discriminating between CIR-driven variability and a binary-induced perturbation can be greatly aided by spatially resolving the innermost wind structures. More broadly, direct spatial information is essential to constrain the geometry, scale, and kinematics of any large-scale structure embedded in the outflow. The potential of long-baseline interferometry to probe large-scale wind asymmetries associated with CIRs was already highlighted by \citet{2002Dessart}, who presented theoretical predictions for spectrally dispersed interferometric observables of rotating, spiral wind structures. Notably, while prior radio observations of WR~6 could only resolve the wind down to $\sim 0.2^{\prime\prime}$ \citep{2000Contreras}, near-infrared interferometry accesses spatial scales of $\sim 0.2$\,mas, achieving a thousand-fold improvement in resolution. In the framework of \citet{2002Dessart}, CIRs produce characteristic, time-dependent signatures in differential visibility and fringe phase, analogous to the S-shaped patterns seen in spectroscopy, and remain detectable even with realistic variations in baseline length and orientation. These results established interferometry as a powerful, complementary diagnostic to spectroscopy for constraining the geometry and dynamics of structured hot-star winds.

Recent advances in optical/near-infrared long-baseline interferometry have now brought this prospect within reach observationally. \citet{2024Deshmukh} demonstrated that modern facilities can partially resolve WR wind emission line regions at sub-au scales and potentially constrain their geometry. Given the expected large-scale, rotating structures associated with CIRs, WR~6 is an ideal target to test these predictions and to directly probe the morphology of its inner wind.

In this paper, we present multi-epoch interferometric observations of WR~6 obtained with the GRAVITY instrument, along with complementary spectroscopic data. Section\,\ref{sec:obs} describes the observations and data reduction. In Section\,\ref{sec:pmoired}, we analyze the GRAVITY data and investigate the spatial structure of the wind. Finally, in Sections\,\ref{sec:discussion} and \ref{sec:conc}, we compare our findings to recent, hydrodynamically-consistent atmosphere modeling of WR~6, assess the contemporaneous spectroscopic modeling, and discuss the implications of our results in the context of CIRs versus binary-induced scenarios.

\section{Observations and data reduction}
\label{sec:obs}

\subsection{Interferometric data}
\begin{figure}
    \centering
    \includegraphics[trim={0cm 0cm 0cm 0cm}, clip,width=0.9\linewidth]{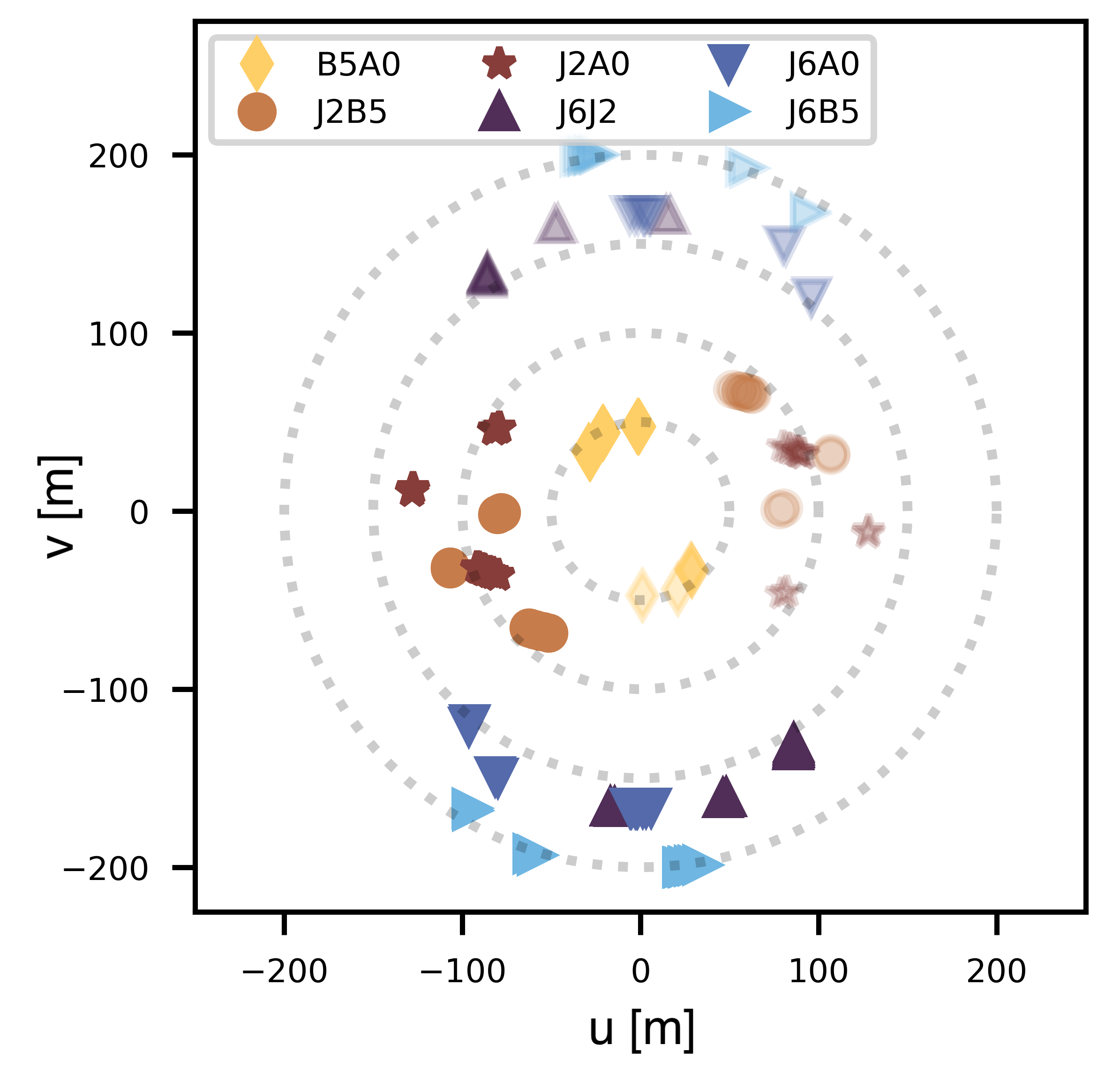}
    \caption{Combined VLTI $uv$-coverage across all six epochs. The different telescope baselines are shown with different markers and colors.}
    \label{fig:uv}
\end{figure}

GRAVITY is a $K$-band instrument at the Very Large Telescope Interferometer (VLTI) at the European Southern Observatory (ESO) in Paranal, Chile \citep{2017GRAVITY}. It offers the powerful capability of obtaining spectro-interferometric observations at high angular resolution. Given that WR~6 is a bright target ($K=5.89$\,mag), we used the high spectral resolution mode ($R\sim4,134$) of GRAVITY over the $1.97-2.40\,\mu\mathrm{m}$ wavelength range. 
The Auxiliary Telescopes (ATs) were used in the extended configuration A0-B5-J2-J6, with baselines up to 202 m, reaching an angular resolution of $\approx1\,$milliarcsecond (mas) in the $K$-band.
We selected HD 49372, a K0III giant with similar magnitude ($K=5.81$\,mag), as the calibrator star from the SearchCal catalog \citep{2011Bonneau}. A total of six epochs were obtained between October 2025 and April 2026, each comprising a 1-hour science-calibrator (SCI-CAL) concatenation. The VLTI $uv$-coverage for all six epochs is shown in Figure\,\ref{fig:uv}. A journal of observations is given in Table\,\ref{tab:obs}, along with the variability phase for each epoch using the 3.76-d period from \citet{1980Firmani}.

   \begin{table}[ht]
      \caption[]{Journal of VLTI/GRAVITY observations of WR~6.}
         \label{tab:obs}
         \renewcommand{\arraystretch}{1.15}
\begin{tabular*}{\columnwidth}{@{\extracolsep{\fill}} cccccc }
\hline
\begin{tabular}{@{}c@{}} Epoch  \end{tabular} 
& \begin{tabular}{@{}c@{}} BJD \end{tabular}  
& \begin{tabular}{@{}c@{}} Seeing ["] \end{tabular} 
& \begin{tabular}{@{}c@{}} $\tau_0$ [ms] \end{tabular} 
& \begin{tabular}{@{}c@{}} $\Delta\phi$ \end{tabular}
& \begin{tabular}{@{}c@{}} Grade \\ \end{tabular} \\

\hline

1 & 60961.285 & 0.70 & 6.49 & 0* & A \\

2 & 60962.288 & 0.64 & 5.19 & 0.27 & A \\

3 & 60963.286 & 0.54 & 7.86 & 0.53 & A \\

4 & 61007.154 & 0.50 & 3.91 & 0.18 & B \\

5 & 61099.064 & 0.48 & 9.18 & 0.59 & A \\

6 & 61138.078 & 0.82 & 6.54 & 0.96 & A \\

\hline

\end{tabular*}
\tablefoot{Listed in the six columns are the epoch numbers, Barycentric Julian Dates (BJDs) of start of observation, dimm seeing during the science target observation, coherence time, variability phase difference ($\Delta\phi$) with respect to the reference phase, and the grade of observation assigned by ESO, respectively. The asterisk represents the reference phase of zero assigned to epoch 1.}
   \end{table}

All observations were reduced and calibrated using the GRAVITY data reduction pipeline, version 1.9.6 \citep{2014Lapeyrere} with the ESO Data Processing System \citep{2024Freudling}. To correct for telluric lines in the $K$-band spectrum, we used the \texttt{molecfit} tool \citep{2015SmetteMolecfit} implemented as the \texttt{tellcorr} function in the interferometric analysis software PMOIRED \citep{2022Merand}. The resulting measured quantities obtained were the visibility amplitude (|V|), closure phase (CP), differential phase (DP), and normalized spectrum (NFLUX).

\subsection{Spectroscopic data}

Spectroscopic data for WR~6 were obtained from two sources.
First, echelle spectra (${R\sim53,000}$) were obtained with the Network of Robotic Echelle Spectrographs (NRES) attached to the 1-m telescopes of the Las Cumbres Observatory (LCO) Global Telescope network \citep{2013PASP..125.1031B}, including at the Wise Observatory and the Cerro Tololo Interamerican Observatory. The NRES spectra were processed with the NRES pipeline. NRES data can be retrieved from the LCO science archive\footnote{\url{https://archive.lco.global/}}. Second, we obtained additional echelle spectra (${R\sim10,000}$) with a 0.5-m telescope on the campus of Embry-Riddle Aeronautical University in Prescott, Arizona, USA. The spectrograph is a commercially-available eShel spectrograph from Shelyak Instruments. The data were recorded with an SBIG ST-10X CCD and then processed with the Demetra software maintained by Shelyak Instruments. This reduction process includes bias, dark, and flat corrections along with a wavelength solution based on contemporaneous ThAr lamp spectra.

\section{Geometric modeling of GRAVITY data}
\label{sec:pmoired}

In this section, we present a general geometric modeling approach for the full GRAVITY wavelength range, followed by the detailed analysis of specific spectral features seen in the data. We use the PMOIRED software as our modeling tool, which uses the Levenberg-Marquardt \citep[LM,][]{1963Marquardt} $\chi^2$ minimization for gradient descent in order to find best fit parameters \citep{2022Merand}. Prior grids can be defined for each free parameter to explore the desired parameter space. This is followed by bootstrapping to estimate uncertainties \citep{1979Efron}. This two-step process is generally robust for relatively simple geometric models with $\lesssim5$ free parameters. 

For larger numbers of free parameters, this approach becomes increasingly impractical. Defining a prior grid fine enough to adequately sample the parameter space scales poorly with dimensionality and becomes computationally expensive. More fundamentally, as the model complexity increases, the $\chi^2$ surface can become highly structured, with multiple local minima and strong degeneracies between parameters. The LM algorithm is a local optimizer by construction, and is therefore susceptible to converging on a local rather than the global minimum, with the outcome often sensitive to the choice of starting point or prior grid. Bootstrapping the fit further compounds the computational cost, since it requires repeating the minimization many times over resampled datasets.

To robustly fit models with a larger number of free parameters, we instead adopt a nested sampling approach \citep{2004Skilling,2020Speagle} implemented within PMOIRED. Unlike the LM algorithm, nested sampling explores the full prior volume rather than descending from a single starting point, making it far less susceptible to becoming trapped in local minima and better suited to sampling complex, multi-modal, or degenerate likelihood surfaces. As a byproduct of the algorithm, it also directly yields the Bayesian evidence, and naturally produces full posterior distributions for each parameter, from which robust uncertainty estimates can be derived without the need for separate bootstrapping.

\subsection{Global properties and companion search with visibilities and closure phases}

\label{sec:v-cp}

\begin{figure}
    \centering
    \includegraphics[trim={0cm 0cm 0cm 0cm}, clip,width=\linewidth]{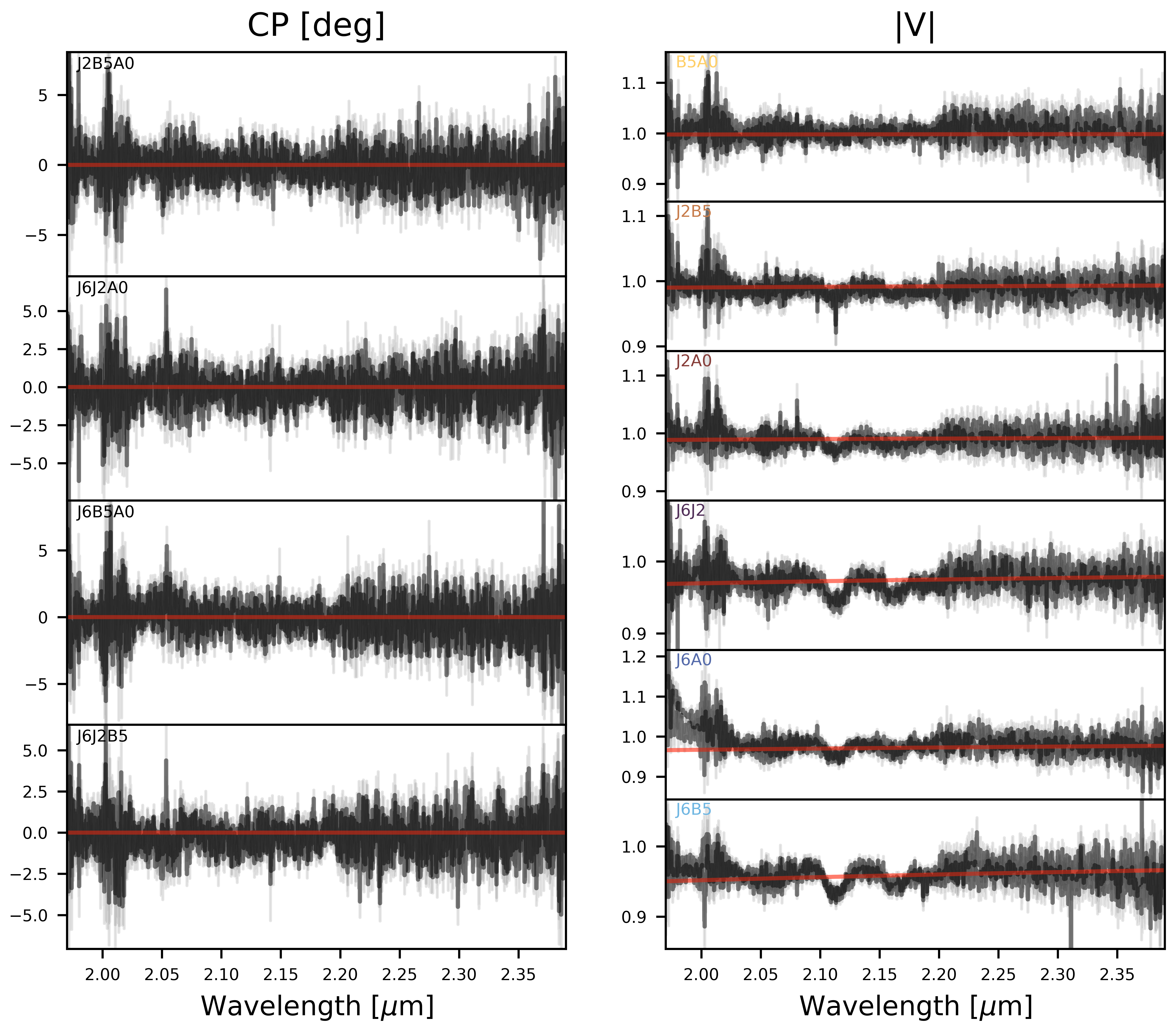}
    \caption{Visibility (|V|) and closure phase (CP) data from the third epoch of GRAVITY observations of WR~6. The data (black) and best-fit uniform disk model (red) are shown as a function of wavelength. The sub-panels correspond to different closure triangles for CP, and different baselines for |V|, which are mentioned in the top right corner in each sub-panel. The |V| sub-panels are ordered, top to bottom, from the shortest to the longest baseline.}
    \label{fig:v2cp}
\end{figure}

For all six epochs of WR~6, the data largely resemble an almost unresolved and symmetric point source (|V| $\approx$ 1 and CP = 0 for all baselines and closure triangles, respectively). We therefore attempt to fit a circular uniform disk model to the data using gradient descent, allowing the diameter to vary. The best-fit diameter values for all epochs are in the range of 0.40--0.49 mas, consistent with being only marginally resolved. Figure\,\ref{fig:v2cp} shows the |V| and CP data from the third epoch, which was modeled as a uniform disk with a diameter of $0.41\pm0.01\,{\rm mas}$ (errors shown hereafter only represent statistical errors).

For each epoch, we also perform a companion search using the grid search algorithm of \citet{2015Gallenne} implemented in PMOIRED. We add a second, unresolved uniform disk of diameter 0.001 mas to the model. With the WR fixed at the origin, the position of the companion is varied over a large grid in the East-North sky-plane spanning $\Delta E = -100\,\,{\rm to}\,100$ mas and $\Delta N = -100\,\,{\rm to}\,100$ mas with steps of 1 mas in both. The flux of the WR is fixed to 1, while that of the companion is allowed to vary. Given the model is still relatively simple, we use gradient descent to find the best solution. On running the grid search, we do not obtain a significant detection of a wide companion in any of the six epochs. 
To quantify the sensitivity of our data, we compute a 3$\sigma$ detection limit on the companion flux for each epoch using the same grid. Across epochs, this limit ranges between 2--3\% of the WR flux, with 95\% of the grid points falling below this value.
For a main sequence companion, this range corresponds to a spectral type of B2.5~V ($M\approx6\,M_\odot$), derived using photometric estimates \citep{2013Pecaut,2019Hamann,2024Deshmukh}. The possible $3-5\,M_\odot$ companion suggested by \citep{2019SchmutzKoenigsbergerWR6} and \citep{2020Koenigsberger} at an orbital period of 3.76 days, if present, would have an angular separation $< 0.1\,$mas, well below the spatial resolution of 1 mas achievable with the |V| and CP observables.

\subsection{Including differential phases and the normalized spectrum}
\label{sec:dphi}

While we can broadly fit the |V|--CP data with a simple uniform disk, there are certain features in |V| that deviate from the model. These features are best visible in the 2.1--2.2 $\mu\mathrm{m}$ range of the bottom three panels of |V| in Figure\,\ref{fig:v2cp}. Such dip-like features in WR data are characteristic of emission at those wavelengths coming from spatially resolved regions \citep{2024Deshmukh}. To investigate this further, we include NFLUX data in the modeling process. Additionally, the DP data from GRAVITY can capture wavelength-dependent spatial asymmetry in the target at sub-mas scales. This is particularly relevant for large-scale structures in WR winds, which contribute to the emission lines and are potentially offset from the stellar photocenter. We therefore include DP to look for deviations from point-symmetry within the spectral features of WR~6. 

Figure\,\ref{fig:zoom} shows the third epoch data in the 2.07--2.24 $\mu\mathrm{m}$ range, NFLUX and DP included (figures for all observations are presented in Appendix\,\ref{sec:app}). The third epoch data are representative of all epochs in terms of the modeling approach and are thus used to describe the general features seen across all epochs. We address key features in the 2.1--2.2 $\mu\mathrm{m}$ range below.

\subsubsection{Qualitative assessment of key features}

The strongest feature in |V| is centered around 2.11 $\mu\mathrm{m}$, roughly spanning the 2.10--2.12 $\mu\mathrm{m}$ range. A corresponding broad emission line can be seen in NFLUX, identified as a blend of two He\,{\sc i} lines $(\lambda_0 = 2.113, 2.114\,\mu\mathrm{m})$. Despite being weaker than the spectral lines around 2.16 and 2.19 $\mu\mathrm{m}$, it shows the strongest dip in |V|, implying that it originates from a more spatially resolved region than the other two. The flat-top shape of the line in the spectrum is indicative of its origin from the optically thin, shell-like region in outer parts of the WR wind. Different parts of the spectral line are emitted from different regions of the wind when projected on the plane of the sky. For example, the blue and red edges of the spectral line originate from parts of the wind along our line-of-sight, coming toward and going away from us, respectively. As a result, the projected size of the line-emitting regions associated with the line-of-sight components is small. On the other hand, the central part of the emission line originates from parts of the wind moving laterally with respect to our line of sight. The projected size in this case is therefore expected to be bigger. 
Qualitatively, this can already be seen from the 2.11 $\mu\mathrm{m}$ feature in |V|. Despite the flat-top shape of the line in NFLUX, the dip in |V| does not have a flat-bottom shape; instead, the central part has the lowest |V|, meaning it is the most spatially resolved part of the line as expected.

\begin{figure*}
    \centering
    \includegraphics[trim={0cm 0cm 0cm 0cm}, clip,width=\linewidth]{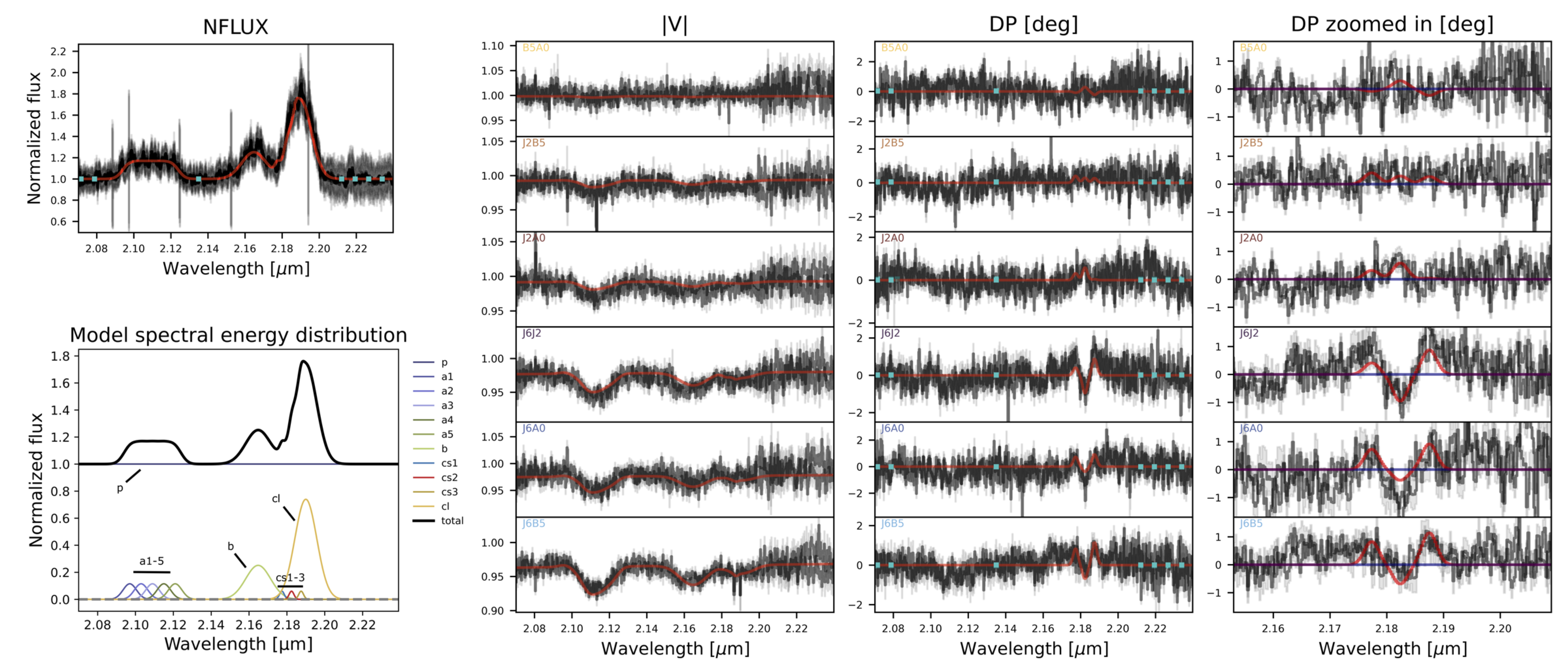}
    \caption{Third epoch GRAVITY data for WR~6 zoomed in to 2.07--2.24 $\mu\mathrm{m}$. Shown here are NFLUX (top of first column), model spectral energy distribution (bottom of first column), |V| (second column), DP (third column), and a zoomed in version of DP (fourth column) data as a function of wavelength. The data are shown in black and error regions are shown in gray, while the best-fit model described in Section\,\ref{sec:dphi} is shown in red. Additionally, the zoomed in DP data are shown along with a model with (red) and without (blue) the \texttt{`cs'} structures included. The |V| and DP sub-panels are ordered, top to bottom, from the shortest to the longest baseline. }
    \label{fig:zoom}
\end{figure*}

The second strongest feature in |V| is observed around 2.16 $\mu\mathrm{m}$. Infrared studies of WN stars identify this spectral region as a blended emission complex involving H\,{\sc i} (Br$\gamma$), He\,{\sc i}, and He\,{\sc ii} transitions \citep[e.g.,][]{1983ApJ...271..221H,1997ApJ...486..420F}.  
Since WR~6 is H-free, the 2.16 $\mu\mathrm{m}$ line is a blend of He\,{\sc i} and He\,{\sc ii} \citep{2019Hamann}.
The dip in |V| is smaller for this line compared to the 2.11 $\mu\mathrm{m}$ He\,{\sc i} line, despite being stronger as seen in NFLUX. This implies the line-forming region for the 2.16 $\mu\mathrm{m}$ line is less spatially resolved than that for the 2.11 $\mu\mathrm{m}$ He\,{\sc i} line, likely due to the significant contribution of He\,{\sc ii} at 2.16 $\mu\mathrm{m}$. The Gaussian-like shape of the spectral line also suggests it forms in a more optically thick region of the wind.

For both the 2.11 and 2.16 $\mu\mathrm{m}$ line regions, there are hints of narrow and sharp features in the DP data at all baselines (Fig.\,\ref{fig:zoom}). These can arise due to the line-forming regions in WR winds being clumpy in nature. Clumps can manifest as photocenter shifts, and therefore non-zero DP values, in the wavelength regions corresponding to these spectral lines. However, the DP features seen around 2.11 and 2.16 $\mu\mathrm{m}$ are weak in strength relative to the noise floor, and we do not attempt to model them.

The strongest spectral line seen in NFLUX is around 2.19 $\mu\mathrm{m}$, identified as a He\,{\sc ii} line $(\lambda_0 = 2.189\,\mu\mathrm{m})$. Given its higher ionization than He\,{\sc i}, the line forms deeper in the wind and is therefore expected to be less spatially resolved in comparison. This can already be seen in |V| data, which shows a negligible dip feature. The global shape of the spectral line as seen in NFLUX can be approximated with a Gaussian profile, which is also expected due to its formation in the optically thick regime.

In Figure\,\ref{fig:zoom}, the DP data around 2.19 $\mu\mathrm{m}$ show significant deviation from zero in the form of multiple features. These are best seen in the bottom three DP panels, corresponding to the three longest baselines. Qualitatively, these features can be explained by discrete line-emitting regions that are offset from the continuum photocenter. 
The global Gaussian profile of the 2.19 $\mu\mathrm{m}$ line results from a relatively symmetric line-forming region, making it difficult to explain off-center features detected via DP. However, large scale structures in the wind can potentially manifest in this manner \citep{2002Dessart}.

\subsubsection{Modeling approach}

The presence of several features at different wavelengths, which are also possibly spatially resolved to different extents, warrants a more robust fitting approach than the simple gradient descent used in Section\,\ref{sec:v-cp}. This is primarily necessitated by a large number of free parameters, some of which are potentially correlated, to ensure a thorough exploration of the subsequent large parameter space. We therefore employ nested sampling and describe the details of the adopted models below.

\subsubsection*{\textit{The continuum}}

We use a uniform disk component \texttt{`p'} to model the $K$-band continuum emission in NFLUX. The component has a normalized flux fixed to 1 by definition, and a diameter allowed to vary over a broad uniform prior range that covers the best-fit diameters of 0.4--0.5 mas found in Section\,\ref{sec:v-cp}.

\subsubsection*{\textit{The 2.11 $\mu\mathrm{m}$ line}}

In order to model the flat-top emission with a varying spatial extent across the 2.10--2.12 $\mu\mathrm{m}$ region, we divide the broad 2.11 $\mu\mathrm{m}$ spectral line into five narrow components namely \texttt{`a1'} through \texttt{`a5'}, such that each narrow component can be assigned its own spatial extent. The amount of components was chosen as a compromise between capturing sufficient details in the data and keeping the number of free parameters within reasonable limits. The spatial intensity distribution of each component in this \texttt{`a'} group is modeled as a simple symmetric Gaussian centered on the star, with its size characterized by the full width at half maximum (fwhm), that can be fit as a free parameter. The emission corresponding to each component is a narrow spectral line with a Gaussian profile, and no contribution to the continuum. The central wavelengths for the five components are spread uniformly over the 2.10--2.12 range and fixed, while the line widths and strengths of all components are tied together, resulting in a flat-topped profile (see Fig.\,\ref{fig:zoom}). This approach allows us to capture the wavelength-dependent spatial extent across the broad line by forcing a discrete decomposition into five components. The shared line width, shared line strength, and individual Gaussian fwhms for the spatial intensity distribution were varied over broad uniform priors. Since the shared line strength and line width can be strongly correlated, we capped the prior for the line width at 9.5 nm to ensure our construction of narrow spectral components stays narrow enough to distinctly observe their different spatial extents encoded in |V|.

\begin{table*}
\centering
\tiny
\caption{Summary of best-fit model parameters for all six epochs as described in Section\,\ref{sec:dphi}. }
\def\arraystretch{1.7}
\begin{tabular}{ccccccccc} \hline \hline
Parameter & Prior & Comments & Epoch 1 & Epoch 2 & Epoch 3 & Epoch 4 & Epoch 5 & Epoch 6  \\ \hline

\texttt{p,f} & -- & {\tiny normalized flux} & 1 & 1 & 1 & 1 & 1 & 1 \\
\texttt{p,ud} [mas] & $\mathcal{U}(0.1, 1.0)$ & {\tiny uniform disk diameter} & $0.448^{+0.001}_{-0.001}$ & $0.402^{+0.001}_{-0.001}$ & $0.369^{+0.001}_{-0.001}$ & $0.482^{+0.001}_{-0.001}$ & $0.411^{+0.001}_{-0.001}$ & $0.463^{+0.001}_{-0.001}$ \\
\hline

\texttt{a1,wl0} [$\mu\mathrm{m}$] & -- & {\tiny spec. line center} & 2.0970 & 2.0970 & 2.0970 & 2.0970 & 2.0970 & 2.0970 \\
\texttt{a2,wl0} [$\mu\mathrm{m}$] & -- & " & 2.103 & 2.103 & 2.103 & 2.103 & 2.103 & 2.103 \\
\texttt{a3,wl0} [$\mu\mathrm{m}$] & -- & " & 2.109 & 2.109 & 2.109 & 2.109 & 2.109 & 2.109 \\
\texttt{a4,wl0} [$\mu\mathrm{m}$] & -- & " & 2.115 & 2.115 & 2.115 & 2.115 & 2.115 & 2.115 \\
\texttt{a5,wl0} [$\mu\mathrm{m}$] & -- & " & 2.121 & 2.121 & 2.121 & 2.121 & 2.121 & 2.121 \\

\texttt{a1-a5,l\_f} & $\mathcal{U}(0.08, 0.20)$ & {\tiny spec. line strength} & $0.109^{+0.003}_{-0.010}$ & $0.133^{+0.002}_{-0.002}$ & $0.115^{+0.001}_{-0.002}$ & $0.115^{+0.001}_{-0.002}$ & $0.118^{+0.002}_{-0.002}$ & $0.120^{+0.002}_{-0.002}$ \\
\texttt{a1-a5,l\_w} [nm] & $\mathcal{U}(5.5, 9.5)$ & {\tiny spec. line width} & $9.46^{+0.04}_{-1.22}$ & $7.35^{+0.08}_{-0.13}$ & $8.31^{+0.07}_{-0.06}$ & $8.62^{+0.10}_{-0.06}$ & $7.92^{+0.12}_{-0.10}$ & $8.22^{+0.10}_{-0.18}$ \\
\texttt{a1-a5,f} & -- & {\tiny continuum flux} & 0 & 0 & 0 & 0 & 0 & 0 \\

\texttt{a1,fwhm} [mas] & $\mathcal{U}(0.01, 1.50)$ & {\tiny spatial extent fwhm} & $0.04^{+0.02}_{-0.03}$ & $0.05^{+0.03}_{-0.04}$ & $0.04^{+0.02}_{-0.03}$ & $0.03^{+0.01}_{-0.02}$ & $0.03^{+0.01}_{-0.02}$ & $0.04^{+0.02}_{-0.03}$ \\
\texttt{a2,fwhm} [mas] & $\mathcal{U}(0.01, 1.50)$ & " & $0.46^{+0.03}_{-0.04}$ & $0.44^{+0.01}_{-0.01}$ & $0.38^{+0.02}_{-0.02}$ & $0.29^{+0.02}_{-0.02}$ & $0.37^{+0.02}_{-0.01}$ & $0.38^{+0.02}_{-0.02}$ \\
\texttt{a3,fwhm} [mas] & $\mathcal{U}(0.01, 1.50)$ & " & $0.80^{+0.04}_{-0.04}$ & $0.60^{+0.01}_{-0.01}$ & $0.74^{+0.01}_{-0.01}$ & $0.53^{+0.02}_{-0.02}$ & $0.68^{+0.01}_{-0.01}$ & $0.49^{+0.02}_{-0.02}$ \\
\texttt{a4,fwhm} [mas] & $\mathcal{U}(0.01, 1.50)$ & " & $0.61^{+0.03}_{-0.02}$ & $0.73^{+0.01}_{-0.01}$ & $0.69^{+0.01}_{-0.01}$ & $0.80^{+0.02}_{-0.02}$ & $0.58^{+0.01}_{-0.01}$ & $0.67^{+0.02}_{-0.02}$ \\
\texttt{a5,fwhm} [mas] & $\mathcal{U}(0.01, 1.50)$ & " & $0.41^{+0.02}_{-0.02}$ & $0.61^{+0.01}_{-0.01}$ & $0.56^{+0.01}_{-0.01}$ & $0.45^{+0.02}_{-0.02}$ & $0.56^{+0.01}_{-0.01}$ & $0.56^{+0.02}_{-0.02}$ \\
\hline

\texttt{b,wl0} [$\mu\mathrm{m}$] & $\mathcal{U}(2.160, 2.175)$ & spec. line center & $2.1639(1)$ & $2.1648(1)$ & $2.1648(1)$ & $2.1654(1)$ & $2.1642(1)$ & $2.1645(1)$ \\
\texttt{b,l\_f} & $\mathcal{U}(0.2, 0.3)$ & {\tiny spec. line strength} & $0.270^{+0.006}_{-0.006}$ & $0.278^{+0.004}_{-0.004}$ & $0.251^{+0.004}_{-0.004}$ & $0.270^{+0.003}_{-0.003}$ & $0.225^{+0.003}_{-0.003}$ & $0.237^{+0.004}_{-0.004}$ \\
\texttt{b,l\_w} [nm] & $\mathcal{U}(10, 20)$ & {\tiny spec. line width} & $14.97^{+0.32}_{-0.34}$ & $14.33^{+0.22}_{-0.23}$ & $15.49^{+0.24}_{-0.24}$ & $15.17^{+0.22}_{-0.22}$ & $14.91^{+0.22}_{-0.24}$ & $15.46^{+0.30}_{-0.30}$ \\
\texttt{b,f} & -- & {\tiny continuum flux} & 0 & 0 & 0 & 0 & 0 & 0 \\
\texttt{b,fwhm} [mas] & $\mathcal{U}(0.1, 1.0)$ & {\tiny spatial extent fwhm} & $0.45^{+0.01}_{-0.01}$ & $0.48^{+0.04}_{-0.05}$ & $0.52^{+0.01}_{-0.01}$ & $0.56^{+0.01}_{-0.01}$ & $0.49^{+0.01}_{-0.01}$ & $0.52^{+0.01}_{-0.01}$ \\
\hline

\texttt{cl,wl0} [$\mu\mathrm{m}$] & $\mathcal{U}(2.187, 2.192)$ & spec. line center & $2.1892(1)$ & $2.1895(1)$ & $2.1900(1)$ & $2.1898(1)$ & $2.1898(1)$ & $2.1899(1)$ \\
\texttt{cl,l\_f} & $\mathcal{U}(0.65, 0.85)$ & {\tiny spec. line strength} & $0.772^{+0.009}_{-0.008}$ & $0.727^{+0.005}_{-0.005}$ & $0.739^{+0.006}_{-0.005}$ & $0.740^{+0.004}_{-0.004}$ & $0.703^{+0.006}_{-0.006}$ & $0.710^{+0.006}_{-0.006}$ \\
\texttt{cl,l\_w} [nm] & $\mathcal{U}(10, 20)$ & {\tiny spec. line width} & $14.34^{+0.13}_{-0.12}$ & $15.15^{+0.12}_{-0.11}$ & $13.70^{+0.06}_{-0.12}$ & $15.15^{+0.08}_{-0.09}$ & $14.58^{+0.09}_{-0.09}$ & $14.92^{+0.08}_{-0.10}$ \\
\texttt{cl,f} & -- & {\tiny continuum flux} & 0 & 0 & 0 & 0 & 0 & 0 \\
\texttt{cl,fwhm} [mas] & $\mathcal{U}(0.1, 1.0)$ & {\tiny spatial extent fwhm} & $0.24^{+0.01}_{-0.01}$ & $0.31^{+0.01}_{-0.01}$ & $0.30^{+0.01}_{-0.01}$ & $0.41^{+0.01}_{-0.01}$ & $0.30^{+0.01}_{-0.01}$ & $0.33^{+0.01}_{-0.01}$ \\

\hline

\texttt{cs\_all,l\_f} & -- & {\tiny spec. line strength} & 0.06 & 0.06 & 0.06 & 0.06 & 0.06 & 0.06 \\
\texttt{cs\_all,l\_w} [nm] & -- & {\tiny spec. line width} & 3 & 3 & 3 & 3 & 3 & 3 \\
\texttt{cs\_all,ud} [mas] & -- & {\tiny pt. source diameter} & 0.001 & 0.001 & 0.001 & 0.001 & 0.001 & 0.001  \\
\texttt{cs1,wl0} [$\mu\mathrm{m}$] & variable & spec. line center & $2.1939(1)$ & $2.1866(1)$ & $2.1774(2)$ & $2.1864(2)$ & $2.1842(1)$ & $2.1860(1)$ \\
\texttt{cs1,x} [mas] & $\mathcal{U}(-1.5,1.5)$ & offset east of origin & $0.034^{+0.017}_{-0.016}$ & $0.361^{+0.017}_{-0.017}$ & $0.048^{+0.012}_{-0.011}$ & $0.208^{+0.017}_{-0.017}$ & $0.206^{+0.011}_{-0.011}$ & $0.320^{+0.016}_{-0.016}$ \\
\texttt{cs1,y} [mas] & $\mathcal{U}(-1.5,1.5)$ & offset north of origin & $0.105^{+0.008}_{-0.009}$ & $0.046^{+0.008}_{-0.008}$ & $0.103^{+0.006}_{-0.006}$ & $0.034^{+0.007}_{-0.007}$ & $-0.046^{+0.008}_{-0.008}$ & $-0.176^{+0.011}_{-0.011}$ \\

\texttt{cs2,wl0} [$\mu\mathrm{m}$] & variable & spec. line center & -- & $2.1954(1)$ & $2.1825(1)$ & $2.1957(1)$ & $2.1924(1)$ & $2.1953(1)$ \\
\texttt{cs2,x} [mas] & $\mathcal{U}(-1.5,1.5)$ & offset east of origin & -- & $0.046^{+0.017}_{-0.017}$ & $0.209^{+0.012}_{-0.012}$ & $0.014^{+0.019}_{-0.018}$ & $-0.151^{+0.012}_{-0.012}$ & $0.022^{+0.017}_{-0.017}$ \\
\texttt{cs2,y} [mas] & $\mathcal{U}(-1.5,1.5)$ & offset north of origin & -- & $0.202^{+0.009}_{-0.009}$ & $-0.070^{+0.007}_{-0.007}$ & $0.230^{+0.007}_{-0.007}$ & $0.193^{+0.009}_{-0.009}$ & $0.185^{+0.012}_{-0.012}$ \\

\texttt{cs3,wl0} [$\mu\mathrm{m}$] & variable & spec. line center & -- & -- & $2.1875(1)$ & -- & -- & -- \\
\texttt{cs3,x} [mas] & $\mathcal{U}(-1.5,1.5)$ & offset east of origin & -- & -- & $-0.056^{+0.016}_{-0.016}$ & -- & -- & -- \\
\texttt{cs3,y} [mas] & $\mathcal{U}(-1.5,1.5)$ & offset north of origin & -- & -- & $0.201^{+0.009}_{-0.009}$ & -- & -- & -- \\

\hline
\end{tabular}
\label{tab:full-params}
\end{table*}

\subsubsection*{\textit{The 2.16 $\mu\mathrm{m}$ line}}

Given the the 2.16 $\mu\mathrm{m}$ is likely dominated by He\,{\sc ii} emission from the optically thick region of the wind, we model it using a component \texttt{`b'} which has a Gaussian spatial intensity distribution, and a Gaussian spectral emission only contributing to the spectral line and not to the continuum. All parameters barring the continuum flux are allowed to vary over broad uniform priors.

\subsubsection*{\textit{The 2.19 $\mu\mathrm{m}$ line}}

Similar to the 2.16 $\mu\mathrm{m}$ line, we model this He\,{\sc ii} feature as component \texttt{`cl'}, a single Gaussian spectral profile centered around 2.19 $\mu\mathrm{m}$ with a Gaussian spatial intensity distribution and no continuum emission. 

The small features observed in DP data in this wavelength region are suspected large scale structures that contribute to the emission but are offset from the photocenter. However, the global shape of the 2.19 $\mu\mathrm{m}$ is largely Gaussian, and no distinct smaller peaks are seen in NFLUX data. We adopt a simple approach to model this combined DP-NFLUX behavior. 

For the third epoch, as shown in Figure\,\ref{fig:zoom}, we add three off-center components namely \texttt{`cs1, cs2, cs3'} based on visual inspection of the DP data. Each component is modeled as a point source emitting only a spectral line and no continuum flux. In reality, large scale wind structures can have complicated geometries \citep[e.g. spirals,][]{2002Dessart}, and constraining them is beyond the capabilities of current GRAVITY data. We allow the positions of all three features to vary across a uniform prior in $\Delta E$ and $\Delta N$, ranging from -1.5 to 1.5 mas in both directions. Their spectral line wavelengths are also varied across uniform priors chosen based on visual inspection of the DP data. 

We attempted several fits allowing for the spectral line width and strength for each component to vary, but their correlation across the \texttt{`cs'} components as well as with the \texttt{`cl'} component often led to relatively poor fits to the DP features. Moreover, the line-strength for the \texttt{`cs'} features is not directly constrained from NFLUX or DP alone. A degeneracy between the line-strength and the offset from the photocenter leads to multiple solutions producing the same DP signal, e.g. a weaker line with a larger photocenter offset, or a stronger line with a smaller photocenter offset. As a simple, first order approach, we therefore fix the line-strength for all such features across all epochs to 0.06 in normalized flux units. We also fix their line width to 3 nm. As a result, the photocenter offsets found for these features are tied to the line-strengths and are to be taken cautiously. Nevertheless, the orientation of the offsets is reliable despite this degeneracy.

\begin{figure*}[htbp]
    \centering
    \includegraphics[trim={0cm 0cm 0cm 0cm}, clip,width=0.95\linewidth]{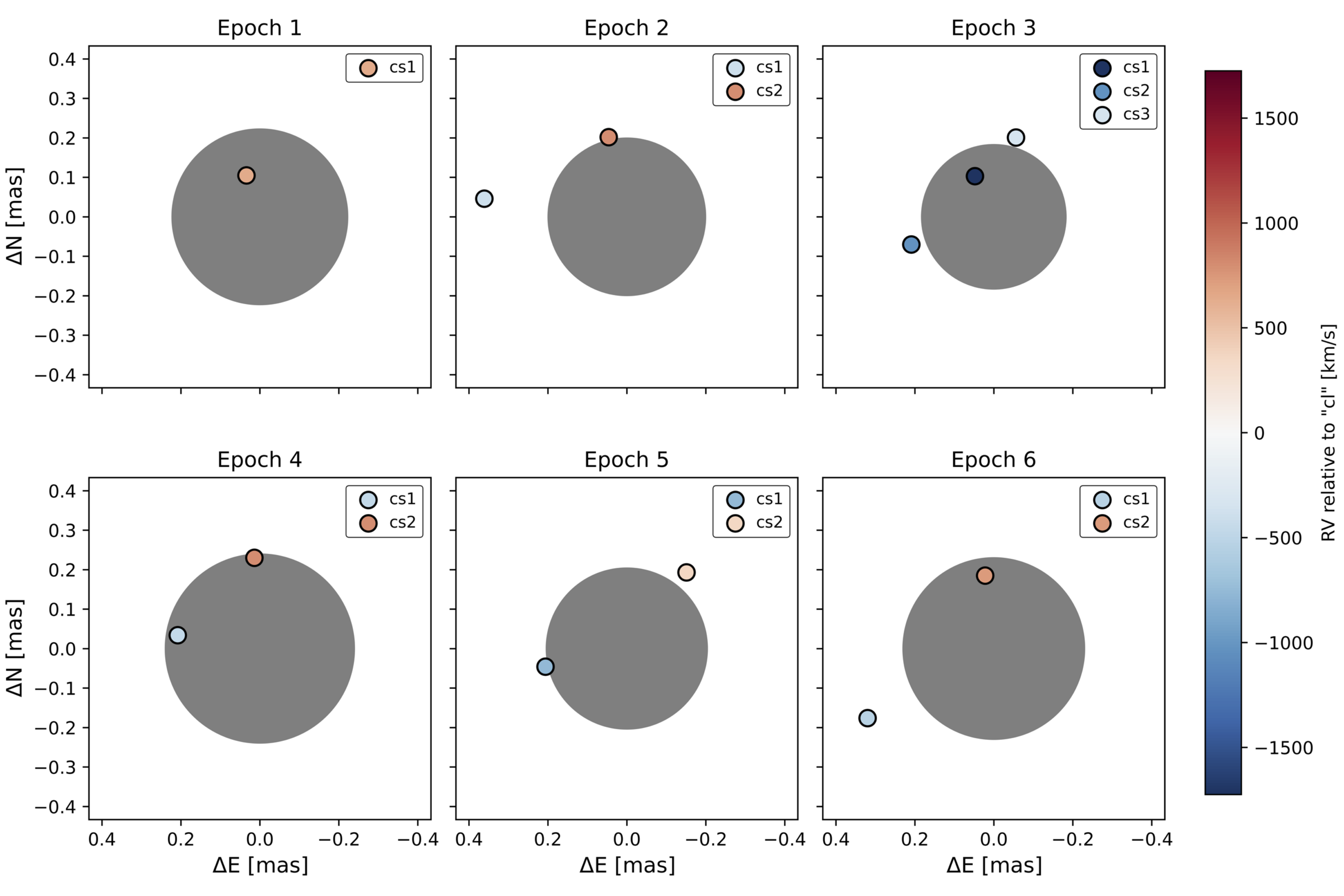}
    \caption{ A visual representation of the best-fit models on the sky-plane for all six epochs of GRAVITY data around the 2.19 $\mu\mathrm{m}$ wavelength region as described in Section\,\ref{sec:dphi}. Each panel shows the sharp continuum disk with diameters given in Table\,\ref{tab:full-params} in gray, along with the DP features modeled using off-center point sources, which are colored by their radial velocity (RV) relative to the central wavelength of the He\,{\sc ii} line (2.189 $\mu\mathrm{m}$).}
    \label{fig:full}
\end{figure*}

\subsubsection*{\textit{Implementation and results}}

In order to weigh the DP data more in the fitting process, we multiplied the DP errors by 0.5 and NFLUX errors by 3. Error bars for uncertainties on the GRAVITY observables are computed using the frame-to-frame scatter in a given exposure, and hence do not take into account systematic uncertainties. Since GRAVITY does not have proper flat fielding \citep[see][section 3.1]{2014Lapeyrere}, the spectra suffer from systematics, which justifies decreasing their weight in the fit by increasing their uncertainties. This allows us to model the DP features around 2.19 $\mu\mathrm{m}$ while assuming a simple Gaussian spectral profile for the strong \texttt{cl} feature. With all components included and uniform priors assigned to free parameters, we performed nested sampling using the \texttt{dynesty} package \citep{2020Speagle}. We used $N_{\rm live}=1000$ live points to sample the posterior in order to maintain a reasonable balance between sampling resolution and computational cost. Bounding of the prior volume was performed using the `multi' method. This method constructs a set of overlapping multi-ellipsoids to both bound the region of higher likelihood and efficiently whiten new draws from the restricted prior as sampling progresses. New live points were drawn using the `rslice' (random slice sampling) method, which is well-suited to problems with a moderate-to-large number of dimensions and can efficiently sample within complex, non-trivial bounding regions. 
We used the resulting unweighted posterior samples to characterize the best-fit parameters and their associated uncertainties. For each free parameter, we report the median of the marginalized posterior distribution, with uncertainties given by the 16th and 84th percentiles (corresponding to the equivalent of a 1$\sigma$ credible interval for a Gaussian distribution). A full summary of the results is provided in Table~\ref{tab:full-params}, while an illustration of the \texttt{`cs'} features on the sky plane for all epochs is shown in Figure\,\ref{fig:full}.

Across all epochs: (i) the best-fit continuum uniform disk diameters range from 0.37--0.48 mas, consistent with the |V|--CP fits from Section\,\ref{sec:v-cp}; (ii) the spatial extent of the 2.11 $\mu\mathrm{m}$ feature typically peaks for the \texttt{a3-a4} components that correspond to emission around 2.109--2.115 $\mu\mathrm{m}$ as expected, while it decreases toward redder and especially bluer wavelengths; (iii) the 2.16 $\mu\mathrm{m}$ feature is barely spatially resolved, and the best-fit parameters show a modest spread, likely due to the line not being a perfect Gaussian; (iv) the dominant 2.19 $\mu\mathrm{m}$ spectral line shows similar behavior to the 2.16 $\mu\mathrm{m}$ line, with an even smaller spatial extent in comparison.

The number of \texttt{cs} features varies from epoch to epoch, ranging from one weakly detected feature in the first epoch, to three relatively robust features in the third epoch. Figure\,\ref{fig:zoom} shows the best-fit model for the third epoch, along with a decomposition of all components. A collection of plots for all epochs is provided in Appendix\,\ref{sec:app}.

\section{Discussion}
\label{sec:discussion}

With multi-epoch VLTI/GRAVITY observations, we newly constrain the spatial extents of line-forming regions in the stellar winds of WR~6. The detections shown in Figure~\ref{fig:full} constitute the first direct interferometric detections of large-scale density structures within a WR wind. Their number and positions are epoch dependent. Below we address the atmospheric properties of WR~6 and structures detected in its wind, and present our interpretation of these detections along with their possible significance.

\subsection{Comparison to atmosphere models}

\citet{2026Lefever} presented the very first spectral analysis of WN4b stars utilizing hydrodynamically-consistent atmosphere models calculated with PoWR$^\textsc{hd}$ \citep{2017Sander,2020Sander,2023Sander}. For stars with dense winds, dynamically-consistent atmosphere models are particularly crucial to infer realistic stellar (i.e., quasi-hydrostatic) radii which can be completely cloaked by the dense wind \citep[e.g.,][]{2004HamannGrafenerWN,2023Lefever}. As the sample of \citet{2026Lefever} includes WR\,6, this provides the unique opportunity to compare a tailored atmosphere/wind model with our interferometric measurements.

The study by \citet{2026Lefever} demonstrates that WN4b stars are very compact with the inferred quasi-hydrostatic radius of WR\,6 being only about $0.75\,R_\odot$. The radius describing the extended photosphere, i.e. $R(\tau = 2/3)$ is already more than twice as large with $\approx2\,R_\odot$. Still, at the distance to WR\,6 \citep[1.4 kpc,][]{2023Crowther}, the $\approx0.2\,$mas radius of our uniform disk fitted to the interferometric observations corresponds to about $60\,R_\odot$ or $\log (r/R_\odot) = 1.8$. At this radius, the wind velocity in the PoWR$^\textsc{hd}$ model is only about $30\,\mathrm{km\,s}^{-1}$ lower than the terminal velocity\footnote{The terminal velocity is taken as the outer boundary value of the PoWR$^\textsc{hd}$ model for WR\,6 at $36\,450\,R_\odot$} of $1759\,\mathrm{km\,s}^{-1}$.

\begin{figure}
    \centering
    \includegraphics[trim={1.2cm 0.2cm 0.5cm 0cm}, clip,width=0.9\linewidth]{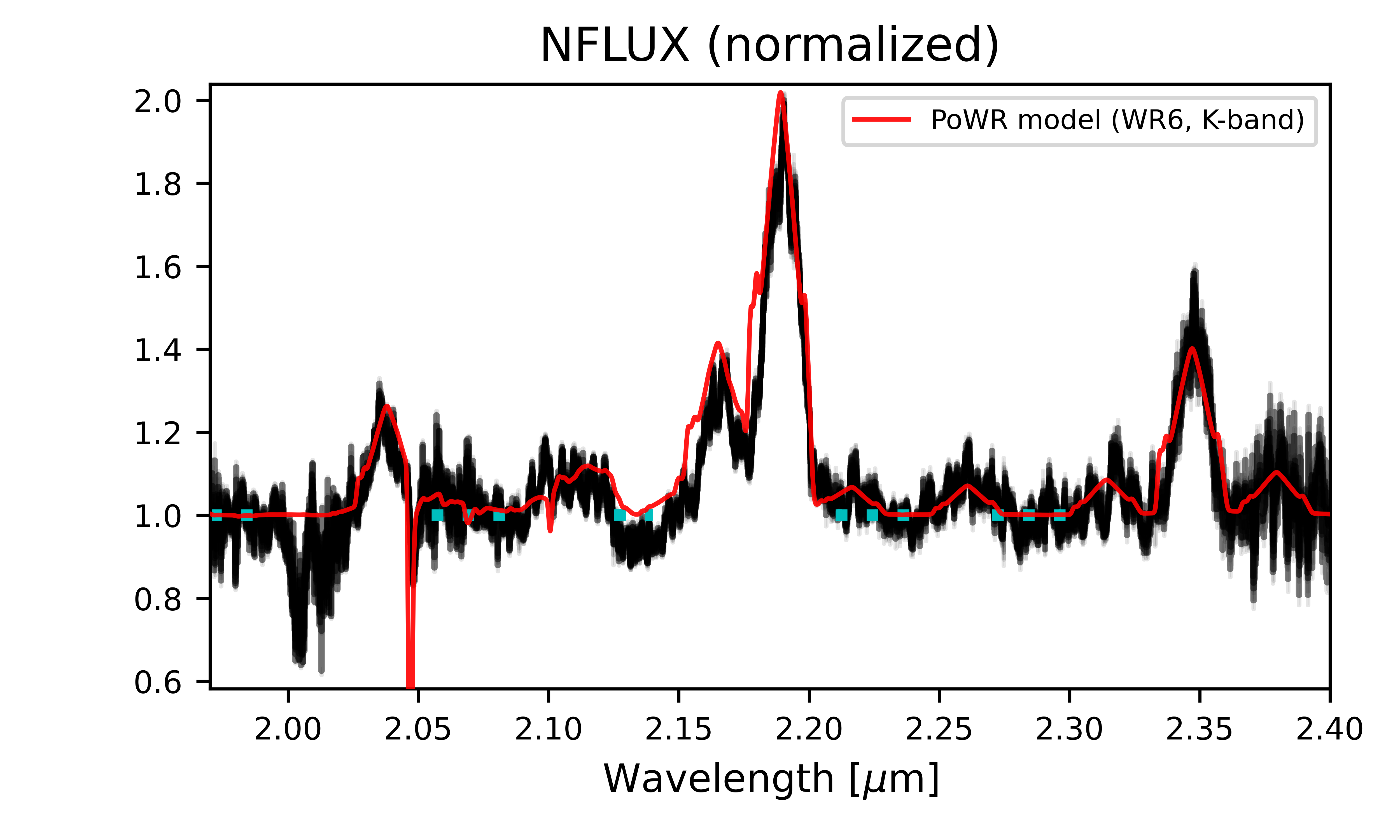}
    \caption{PoWR$^\textsc{hd}$ model from \citet{2026Lefever} for WR~6 (red) shown with the third epoch spectrum from GRAVITY (black).}
    \label{fig:powr}
\end{figure}

\begin{figure}
    \centering
    \includegraphics[trim={0.4cm 0.4cm 0.3cm 0cm}, clip,width=0.9\linewidth]{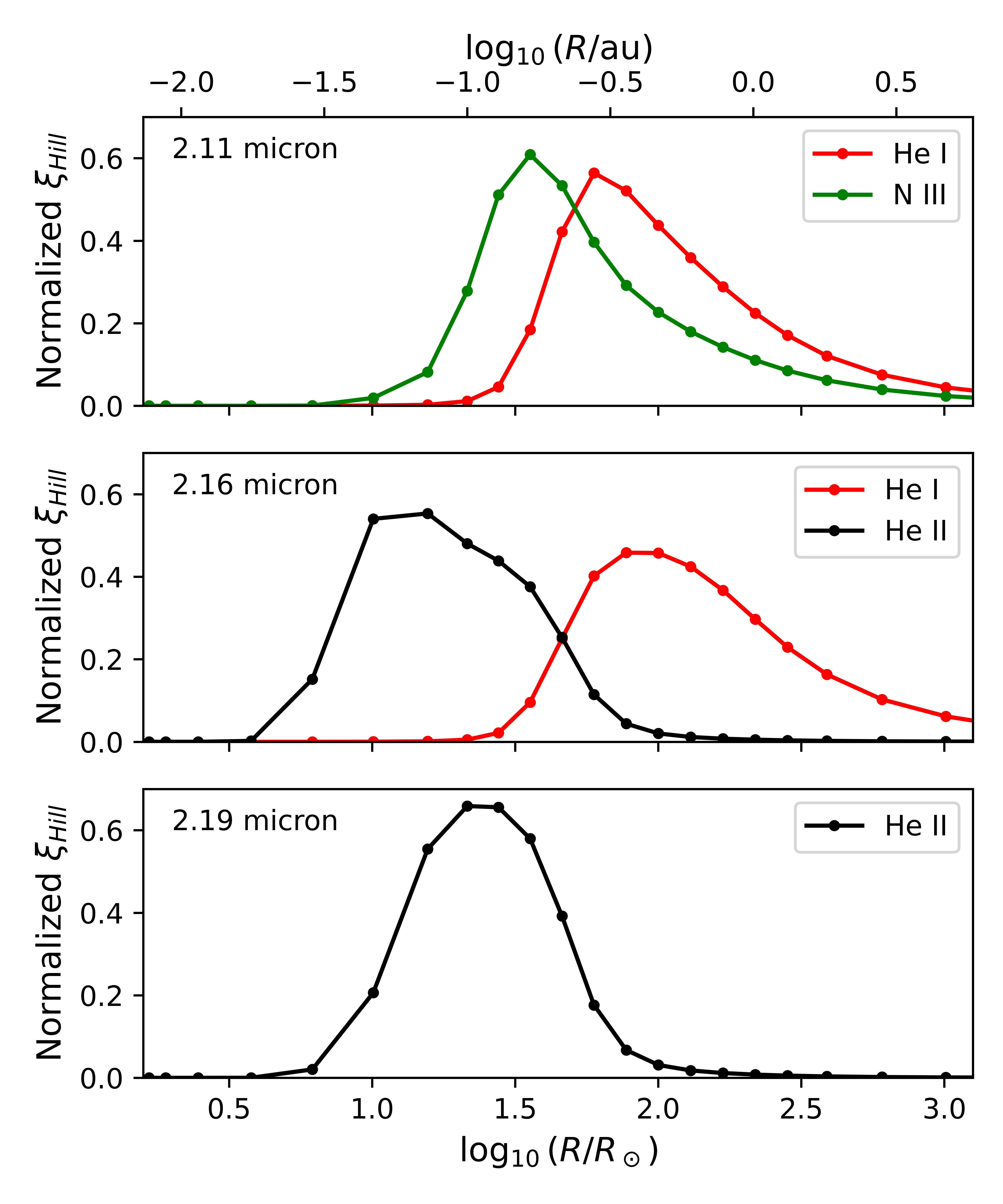}
    \caption{Line formation as a function of radius for selected K-band emission lines as predicted by the PoWR$^\textsc{hd}$ for WR\,6. }
    \label{fig:lfr}
\end{figure}

Although the spectral analysis of WR\,6 in \citet{2026Lefever} did not use any $K$-band observations, the synthetic spectrum from the best-matching model (determined from optical and UV) shows an excellent match with the GRAVITY $K$-band spectrum as illustrated in Figure\,\ref{fig:powr}. In Figure\,\ref{fig:lfr}, we show line formation plots for the prominent emission lines at $2.11$, $2.16$, and $2.19\,\mu$m based on the atmosphere model. These are calculated using the emission measure $\xi_\text{Hill}$ introduced by \citet{1987Hillier} in a study of WR\,6, at that time to study the optical He emission line formation \citep[soon after also for C and N lines in][]{1988Hillier}. When comparing the curves with the reference radius of the uniform disk, it immediately becomes clear that the \ion{He}{ii} lines form further inwards and are thus unresolved, while the \ion{He}{i} lines mostly form at $\log (r/R_\odot) \gtrsim 1.8$ and thus are resolved. Notably, the atmosphere model predicts the $2.11\,\mu$m-line to form a little bit more inward than the $2.16\,\mu$m-line, despite the |V|-signal being stronger for the former one. The reason for this is likely the weaker contribution of \ion{He}{i} to the $2.16\,\mu$m blend. 

Overall, the interferometric measurements convincingly confirm the line formation for the bulk wind motion predicted by the PoWR$^\textsc{hd}$ model for WR\,6.

\subsection{Description of detected structures}

All six epochs of WR~6 observations show at least one DP feature around $2.19\,\mu$m that we model as an off-center point source contributing flux in the form of weak and narrow emission. Analogous narrow emission components that vary over time are ubiquitous in time-series spectroscopic observations of WR stars. However, they are often stochastic and are attributed to over-dense clumps in the WR wind \citep{2008Moffat,2020Chene}, although no spatially resolved counterparts have been reported before. In GRAVITY observations of WR~6, we see between one and three distinct features extending out to several tens of stellar radii. We therefore interpret them as large-scale structures in the winds of WR~6. 

Each epoch exhibits one structure located north of $\Delta N\approx0.1$~mas. With the exception at Epoch 3, all of these detections are redshifted with respect to the \texttt{`cl'} feature. Across the first three epochs taken over three consecutive nights, and assuming \texttt{`cs1'} is the same structure in all of them, its wavelength evolved from redshifted to blueshifted. 
Five out of the six epochs also exhibit at least a second structure, always located in the eastern hemisphere. These structures are all blueshifted. A third structure, appearing uniquely in Epoch 3, is the most blueshifted.

No structures are detected in the southwest quadrant over the six epochs. While the overall morphology appears qualitatively repetitive, the configurations are not identical from one epoch to another. Epoch 4, whose phase falls between those of Epochs 1 and 2, may correspond to an intermediate configuration compatible with the evolution observed during the first three epochs. However, this is not the case for Epochs 5 and 6, whose phases are respectively close to those of Epochs 3 and 1. The positions and Doppler shifts of the detected structures are too different to correspond to the same stable configuration repeating over multiple cycles.

\begin{table}
\caption{Individual detections of discrete off-center features around 2.19 $\mu\mathrm{m}$ grouped by epoch.}
\renewcommand{\arraystretch}{1.15}
\centering
\begin{tabular}{ccccr @{\,$\pm$\,} l}
\hline
Ep. & $\Delta$d (day) & $\Delta\phi$ & ID & \multicolumn{2}{r}{$\xi$ (km~s$^{-1}$)}  \\
\hline

1 & 0 & 0 & \texttt{cs1} &  671&14 \\

\hline
2 & 1.0 & 0.27 & \texttt{cs1} &  --329&14 \\
   & & & \texttt{cs2} &  876&14 \\

\hline
3 & 2.0 & 0.53 & \texttt{cs1} & --1589&28 \\
   & & & \texttt{cs2} & --890&14 \\
   & & & \texttt{cs3} & --205&14 \\

\hline
4 & 45.9 & 0.18 & \texttt{cs1} & --356&28 \\
   & & & \texttt{cs2} & 918&14 \\

\hline
5 & 137.8 & 0.59 & \texttt{cs1} & --657&14 \\
   & & & \texttt{cs2} & 466&14 \\

\hline
6 & 176.8 & 0.95 & \texttt{cs1} & --411&14 \\
   & & & \texttt{cs2} & 863&14 \\
   
\hline
\end{tabular}
\vspace{3pt}
\tablefoot{ Listed for each detection are the time after first epoch ($\Delta$d), variability phase with respect to the first epoch line-of-sight velocity ($\xi$) relative to the rest-wavelength $\lambda_0 = 2.189\,\mathrm{\mu}m$ with uncertainties.}
\label{tab:detections}
\end{table}

\begin{figure}
    \centering
    \includegraphics[width=0.95\linewidth]{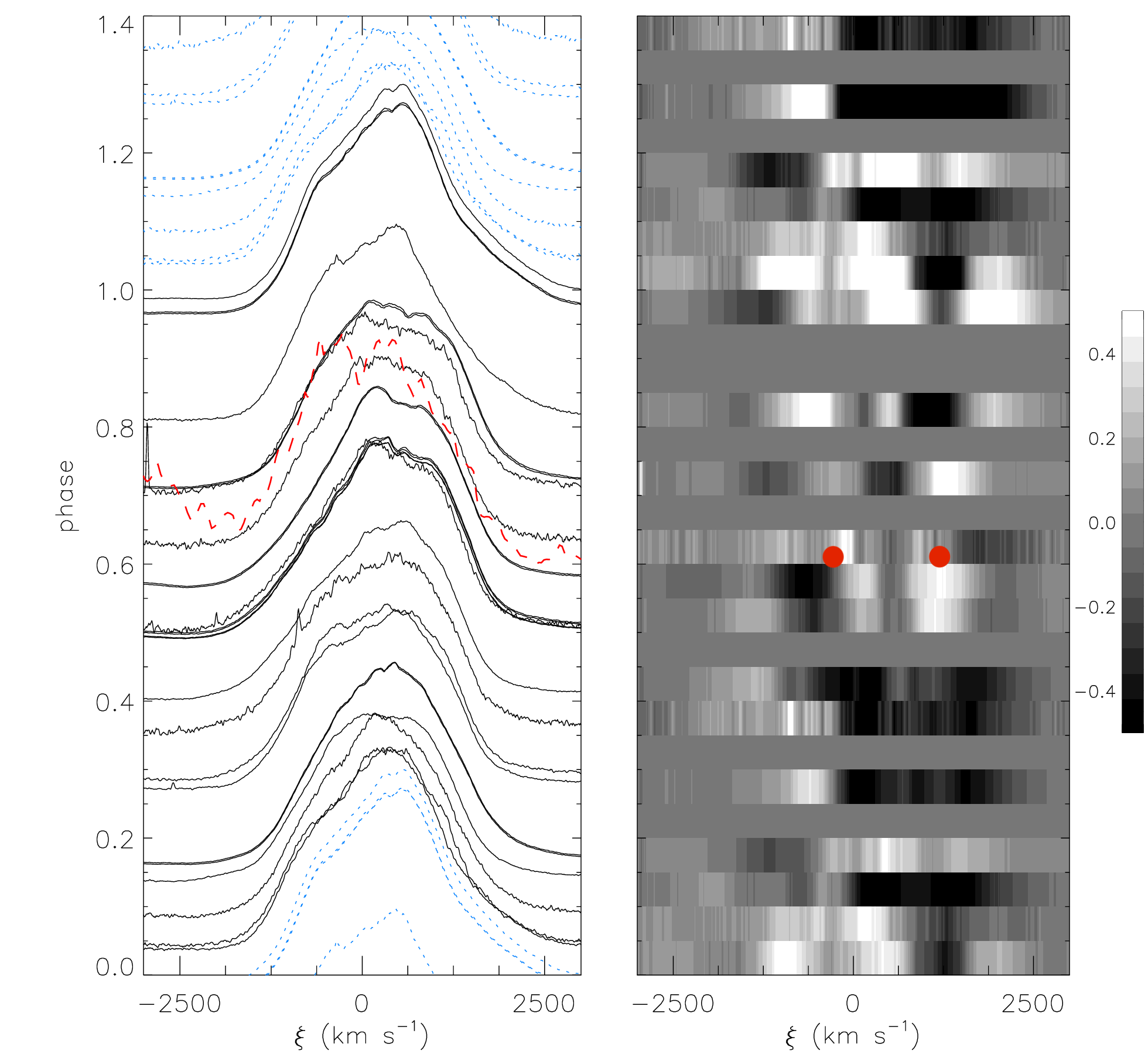}
    \caption{Contemporaneous monitoring of the He{\sc ii}$\lambda$4686 line around the time of Epoch~5. Left: individual spectra folded on the 3.76~d period of WR~6. Black solid lines show the observed spectra, while blue dotted lines correspond to the same spectra shifted by one cycle to illustrate continuity across phases. The red dashed line indicates the 2.19~$\mu$m He{\sc ii} line observed with GRAVITY at Epoch 5. Right: Dynamical spectrum of the residual line-profile variability obtained after subtraction of the mean spectrum. Residuals were averaged within phase bins of width 0.05. At the phase of the GRAVITY observations, two positive residual features (marked with red dots) are visible on opposite sides of the line profile, moving respectively toward more negative and more positive velocities.}
    \label{fig:Spectro}
\end{figure}

\subsection{Contemporaneous spectroscopic monitoring}\label{sec:Spectro}

The line-of-sight velocity ($\xi$) of a structure provides information on its position relative to the observer, but alone does not constrain the direction of rotation. Contemporaneous spectroscopy therefore allows a more robust interpretation of the evolution of the detected structures over time.

Figure~\ref{fig:Spectro} shows the line-profile variability of the He{\sc ii}\,$\lambda$4686 line as a function of phase, assuming a 3.76~d period \citep{1980Firmani}. These spectra were obtained over a two-week interval that includes the night of Epoch 5. At the phase corresponding to the GRAVITY observations, two dominant positive residual structures are visible: one on the blue side of the profile, evolving towards increasingly negative velocities, and another on the red side, evolving towards increasingly positive velocities. Associating the blueshifted residual trace with the blue interferometric structure detected during Epoch 5 shown in Figure\,\ref{fig:full}, and the redshifted residual trace with the corresponding red structure, implies a counter-clockwise rotation of the wind structures on the plane of the sky.

\subsection{Analysis of the first 3 epochs}

The first three epochs were obtained over three consecutive nights, providing an opportunity to study the evolution of the wind structures over half a complete variability cycle.

The two main scenarios proposed to explain the origin of the large-scale wind structures are a wind-companion interaction \citep{2019SchmutzKoenigsbergerWR6} and intrinsic surface perturbations generating CIRs \citep{1997Morel}. In both cases, the underlying principle is similar: a density enhancement is launched into the wind and subsequently carried away while being curved by stellar rotation or orbital motion. We therefore constructed a simple geometric model describing the large-scale structures associated with the features detected in Figure~\ref{fig:full}.

The fundamental parameters can be adopted directly from the literature: the rotational/orbital period \citep[3.76~d,][]{1980Firmani}, the distance to the star \citep[1.4~kpc,][]{2023Crowther}, and the wind velocity \citep[i.e., $v_\infty=1760$~km~s$^{-1}$][]{2026Lefever}. 
 
The position of the detected structures along the line of sight ($\Delta Z$) can be estimated from $\Delta E$, $\Delta N$, and the Doppler shift measured in the DP maps:
\begin{equation}
 \Delta Z=\frac{\sqrt{\Delta E^2+\Delta N^2}}{\arctan{\theta}};~~    \theta=\arccos\left(\frac{\lambda-\lambda_0}{\lambda_0} \frac{c}{v_\infty}\right),
\end{equation}
\noindent where $\lambda_0=2.189~\mu m$ for the He{\sc ii} line.

The model assumes that once launched, the structures are passively advected by the wind while preserving their latitude relative to the rotation/orbital axis. The declination angle of a structure is therefore given by:
\begin{equation}
\delta = \arcsin\left( \frac{\Delta E \sin{i} \cos{\gamma}+\Delta Z \sin{i} \cos{\gamma}+\Delta N \cos{i}}{\sqrt{\Delta E^2 + \Delta Z^2 + \Delta N^2} }\right),
\end{equation}
\noindent where $i$ is the inclination of the rotation/orbital axis relative to the $\Delta E$ axis and $\gamma$ is the angle between the axis and the line of sight.

The only free parameters of the geometric model are therefore $i$, $\gamma$, and the sense of rotation/orbit. The latter can be constrained from the spectroscopic monitoring presented in Section~\ref{sec:Spectro}. If the same structure is observed over the three consecutive nights, it should preserve a constant declination angle ($\delta$), while its rotational/orbital phase should evolve by approximately $360^\circ/3.76~\mathrm{d}\sim96^\circ$ per night.

There are no formal solutions that fulfill all the conditions. In the closest case, when $i=50^\circ$ and $\gamma=10^\circ$, there is a $\sim140^\circ$ difference between the origins of one structure detected in the three epochs. Would this solution be right, the declination angle of the structure would migrate from $100^\circ$, to $50^\circ$, to $15^\circ$ within 3 days, which is not necessarily impossible but incompatible with photometric and spectroscopic monitoring of WR\,6 as they do not suggest such a rapid migration. Figure~\ref{fig:GeoMod} shows the corresponding model. This model depends on the following set of assumptions:

\begin{figure}
    \centering
    \includegraphics[width=0.99\linewidth]{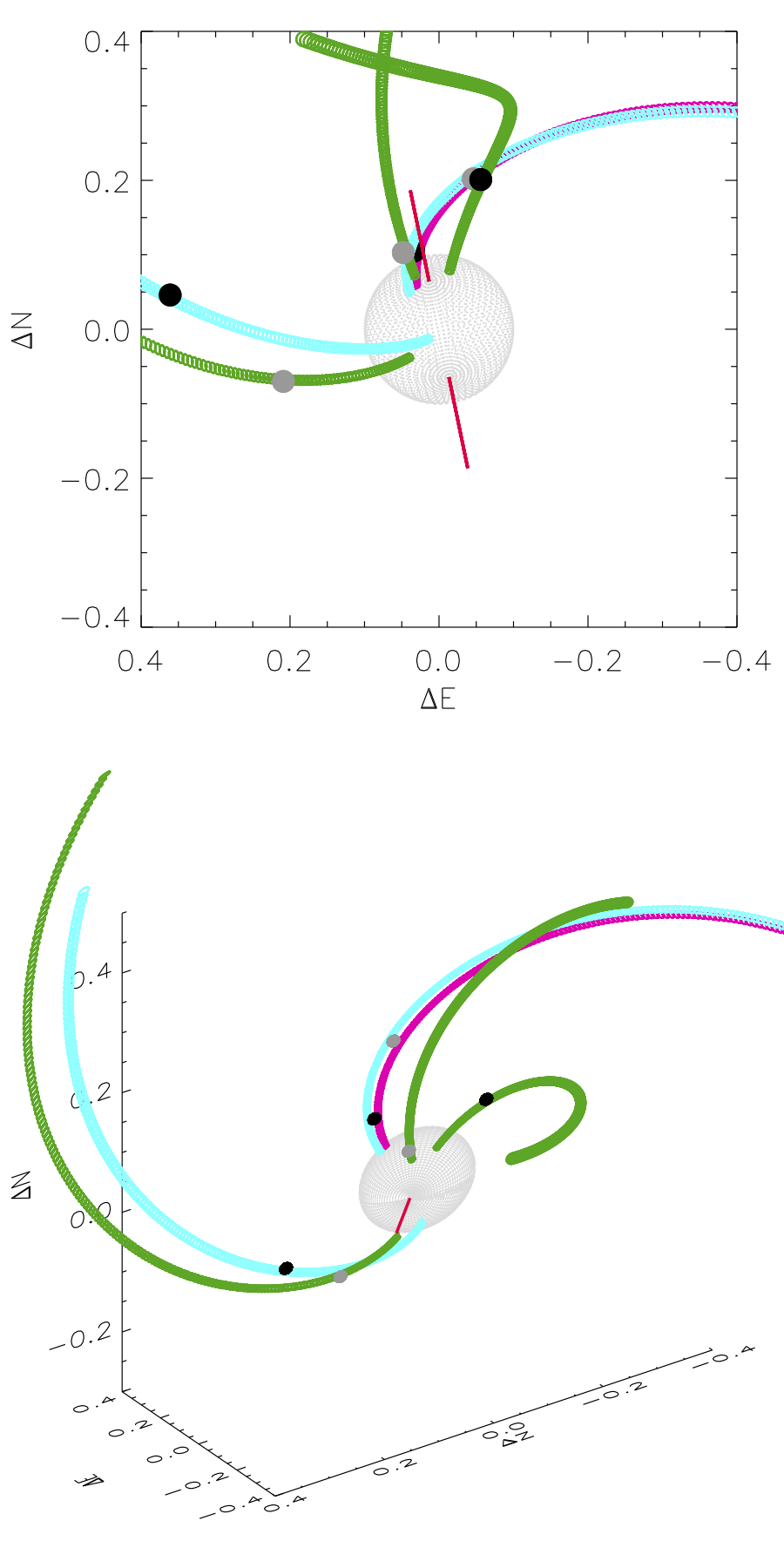}
    \caption{Geometric model of the wind structures detected during the first three epochs, assuming a rotation/orbital axis inclination of $i=50^\circ$ and an angle of $\gamma=10^\circ$ with respect to the line of sight. The purple structure corresponds to Epoch 1, while the blue and green structures correspond to Epoch 2 and Epoch 3, respectively. The gray points mark the detections and the black points identify the structures associated with Epoch 1 under the assumption that they represent the same physical feature observed at successive epochs. The red line indicates the rotation/orbital axis.}
    \label{fig:GeoMod}
\end{figure}

\begin{itemize}
    \item Each detection corresponds to only one structure.
    \item The structures can be detected at all phases, regardless of their density configuration.
    \item We have detected all of the structures.
\end{itemize}
The fact that we cannot link the configuration of the three epochs is an indication that those assumptions cannot all apply at once. Moreover, this approach is also limited by our first order approximation in modeling the possibly complex-shaped structures as point sources in Section\,\ref{sec:dphi}.

To better constrain the angle of rotation or orbit, a better sampled monitoring is necessary. For example, with two observations separated by at least 4.5 hours (0.05 in phase) per night for at least 5 nights within the course of a month would allow us to better follow the evolution of the detections.

\section{Conclusion}
\label{sec:conc}

We present the first interferometric detection of large-scale density structures in the wind of Wolf-Rayet stars. The number of detected structures and their positions are incompatible with a single structure caused by the presence of a low-mass companion. We therefore favor the corotating interaction region (CIR) scenario to explain our observations, while also supporting previous photometric, spectroscopic, and polarimetric observations. Furthermore, our interpretation of the observations shows that the star's rotation is counter-clockwise.

The CIR structures were found mostly in pairs, but they were never found in opposition, as it is usually represented in hydrodynamic models \citep[e.g.,][]{1996Cranmer,2002Dessart}. Their configuration is epoch-dependent, as expected, yet previous studies suggest a time-period of coherence of 40 to 60 days. A high-cadence interferometric campaign in the future will likely provide more insights into the evolution of CIRs.

\begin{acknowledgements}
The authors thank Em Biegler, Ari Chai, William Punches, Jesse Linden, and Hailey Widger for assisting with spectroscopic observations used in this work. This research is based on observations collected at the European Southern Observatory under ESO program ID 116.290L (PI. Deshmukh) and on spectroscopic observations collected at the Las Cumbres Observatory under LCOGT program LTP2025B-001 (PI. Richardson). This research received support from Long term structural funding - Methusalem funding by the Flemish Government (project SOUL, METH/24/012) at KU Leuven. KD and AJF acknowledge support from the ESO Early-Career Scientific Visitor Programme. Based in part on data obtained from the LCO Observatory through program LTP2025B-001. TDP acknowledges support of the Research Foundation - Flanders (FWO) under grant 11P6I24N. NDR is grateful for support from the Cottrell Scholar Award \#CS-CSA-2023-143 sponsored by the Research Corporation for Science Advancement.
RRL and AACS acknowledge funding by the Deutsche Forschungsgemeinschaft (DFG, German Research Foundation) in the form of an Emmy Noether Research Group – Project-ID 445674056 (PI Sander). RRL and AACS further acknowledge support from the Federal Ministry of Education and Research (BMBF) and the Baden-W{\"u}rttemberg Ministry of Science as part of the Excellence Strategy of the German Federal and State Governments. This project was co-funded by the European Union (Project 101183150 - OCEANS).
TS acknowledges support from the Israel Science Foundation (ISF) under grant number 0603225041 and from the European Research Council (ERC) under the European Union’s Horizon 2020 research and innovation program (grant agreement 101164755/METAL).
This research was supported by the Munich Institute for Astro-, Particle and BioPhysics (MIAPbP) which is funded by the Deutsche Forschungsgemeinschaft (DFG, German Research Foundation) under Germany's Excellence Strategy - EXC-2094 - 390783311.

\end{acknowledgements}

\bibliographystyle{aa} 
\bibliography{aanda} 

@ARTICLE{2013PASP..125.1031B,
       author = {{Brown}, T.~M. and {Baliber}, N. and {Bianco}, F.~B. and {Bowman}, M. and {Burleson}, B. and {Conway}, P. and {Crellin}, M. and {Depagne}, {\'E}. and {De Vera}, J. and {Dilday}, B. and {Dragomir}, D. and {Dubberley}, M. and {Eastman}, J.~D. and {Elphick}, M. and {Falarski}, M. and {Foale}, S. and {Ford}, M. and {Fulton}, B.~J. and {Garza}, J. and {Gomez}, E.~L. and {Graham}, M. and {Greene}, R. and {Haldeman}, B. and {Hawkins}, E. and {Haworth}, B. and {Haynes}, R. and {Hidas}, M. and {Hjelstrom}, A.~E. and {Howell}, D.~A. and {Hygelund}, J. and {Lister}, T.~A. and {Lobdill}, R. and {Martinez}, J. and {Mullins}, D.~S. and {Norbury}, M. and {Parrent}, J. and {Paulson}, R. and {Petry}, D.~L. and {Pickles}, A. and {Posner}, V. and {Rosing}, W.~E. and {Ross}, R. and {Sand}, D.~J. and {Saunders}, E.~S. and {Shobbrook}, J. and {Shporer}, A. and {Street}, R.~A. and {Thomas}, D. and {Tsapras}, Y. and {Tufts}, J.~R. and {Valenti}, S. and {Vander Horst}, K. and {Walker}, Z. and {White}, G. and {Willis}, M.},
        title = "{Las Cumbres Observatory Global Telescope Network}",
      journal = {\pasp},
         year = 2013,
        month = sep,
       volume = {125},
       number = {931},
        pages = {1031},
          doi = {10.1086/673168},
archivePrefix = {arXiv},
       eprint = {1305.2437},
 primaryClass = {astro-ph.IM},
       adsurl = {https://ui.adsabs.harvard.edu/abs/2013PASP..125.1031B}
}

@INPROCEEDINGS{2022Merand,
       author = {{M{\'e}rand}, Antoine},
        title = "{Flexible spectro-interferometric modelling of OIFITS data with PMOIRED}",
    booktitle = {Optical and Infrared Interferometry and Imaging VIII},
         year = 2022,
       editor = {{M{\'e}rand}, Antoine and {Sallum}, Stephanie and {Sanchez-Bermudez}, Joel},
       series = {Society of Photo-Optical Instrumentation Engineers (SPIE) Conference Series},
       volume = {12183},
        month = aug,
          eid = {121831N},
        pages = {121831N},
          doi = {10.1117/12.2626700},
archivePrefix = {arXiv},
       eprint = {2207.11047},
 primaryClass = {astro-ph.IM},
       adsurl = {https://ui.adsabs.harvard.edu/abs/2022SPIE12183E..1NM}
}

@ARTICLE{2017GRAVITY,
       author = {{GRAVITY Collaboration} and {Abuter}, R. and {Accardo}, M. and {Amorim}, A. and {Anugu}, N. and {{\'A}vila}, G. and {Azouaoui}, N. and {Benisty}, M. and {Berger}, J.~P. and {Blind}, N. and {Bonnet}, H. and {Bourget}, P. and {Brandner}, W. and {Brast}, R. and {Buron}, A. and {Burtscher}, L. and {Cassaing}, F. and {Chapron}, F. and {Choquet}, {\'E}. and {Cl{\'e}net}, Y. and {Collin}, C. and {Coud{\'e} Du Foresto}, V. and {de Wit}, W. and {de Zeeuw}, P.~T. and {Deen}, C. and {Delplancke-Str{\"o}bele}, F. and {Dembet}, R. and {Derie}, F. and {Dexter}, J. and {Duvert}, G. and {Ebert}, M. and {Eckart}, A. and {Eisenhauer}, F. and {Esselborn}, M. and {F{\'e}dou}, P. and {Finger}, G. and {Garcia}, P. and {Garcia Dabo}, C.~E. and {Garcia Lopez}, R. and {Gendron}, E. and {Genzel}, R. and {Gillessen}, S. and {Gonte}, F. and {Gordo}, P. and {Grould}, M. and {Gr{\"o}zinger}, U. and {Guieu}, S. and {Haguenauer}, P. and {Hans}, O. and {Haubois}, X. and {Haug}, M. and {Haussmann}, F. and {Henning}, Th. and {Hippler}, S. and {Horrobin}, M. and {Huber}, A. and {Hubert}, Z. and {Hubin}, N. and {Hummel}, C.~A. and {Jakob}, G. and {Janssen}, A. and {Jochum}, L. and {Jocou}, L. and {Kaufer}, A. and {Kellner}, S. and {Kendrew}, S. and {Kern}, L. and {Kervella}, P. and {Kiekebusch}, M. and {Klein}, R. and {Kok}, Y. and {Kolb}, J. and {Kulas}, M. and {Lacour}, S. and {Lapeyr{\`e}re}, V. and {Lazareff}, B. and {Le Bouquin}, J.-B. and {L{\`e}na}, P. and {Lenzen}, R. and {L{\'e}v{\^e}que}, S. and {Lippa}, M. and {Magnard}, Y. and {Mehrgan}, L. and {Mellein}, M. and {M{\'e}rand}, A. and {Moreno-Ventas}, J. and {Moulin}, T. and {M{\"u}ller}, E. and {M{\"u}ller}, F. and {Neumann}, U. and {Oberti}, S. and {Ott}, T. and {Pallanca}, L. and {Panduro}, J. and {Pasquini}, L. and {Paumard}, T. and {Percheron}, I. and {Perraut}, K. and {Perrin}, G. and {Pfl{\"u}ger}, A. and {Pfuhl}, O. and {Phan Duc}, T. and {Plewa}, P.~M. and {Popovic}, D. and {Rabien}, S. and {Ram{\'\i}rez}, A. and {Ramos}, J. and {Rau}, C. and {Riquelme}, M. and {Rohloff}, R.-R. and {Rousset}, G. and {Sanchez-Bermudez}, J. and {Scheithauer}, S. and {Sch{\"o}ller}, M. and {Schuhler}, N. and {Spyromilio}, J. and {Straubmeier}, C. and {Sturm}, E. and {Suarez}, M. and {Tristram}, K.~R.~W. and {Ventura}, N. and {Vincent}, F. and {Waisberg}, I. and {Wank}, I. and {Weber}, J. and {Wieprecht}, E. and {Wiest}, M. and {Wiezorrek}, E. and {Wittkowski}, M. and {Woillez}, J. and {Wolff}, B. and {Yazici}, S. and {Ziegler}, D. and {Zins}, G.},
        title = "{First light for GRAVITY: Phase referencing optical interferometry for the Very Large Telescope Interferometer}",
      journal = {\aap},
         year = 2017,
        month = jun,
       volume = {602},
          eid = {A94},
        pages = {A94},
          doi = {10.1051/0004-6361/201730838},
archivePrefix = {arXiv},
       eprint = {1705.02345},
 primaryClass = {astro-ph.IM},
       adsurl = {https://ui.adsabs.harvard.edu/abs/2017A&A...602A..94G}
}

@ARTICLE{2012Langer,
       author = {{Langer}, N.},
        title = "{Presupernova Evolution of Massive Single and Binary Stars}",
      journal = {\araa},
         year = "2012",
        month = "Sep",
       volume = {50},
        pages = {107-164},
          doi = {10.1146/annurev-astro-081811-125534},
archivePrefix = {arXiv},
       eprint = {1206.5443},
 primaryClass = {astro-ph.SR},
       adsurl = {https://ui.adsabs.harvard.edu/abs/2012ARA&A..50..107L}
}

@ARTICLE{2024Deshmukh,
       author = {{Deshmukh}, K. and {Sana}, H. and {M{\'e}rand}, A. and {Bordier}, E. and {Langer}, N. and {Bodensteiner}, J. and {Dsilva}, K. and {Frost}, A.~J. and {Gosset}, E. and {Le Bouquin}, J. -B. and {Lefever}, R.~R. and {Mahy}, L. and {Patrick}, L.~R. and {Reggiani}, M. and {Sander}, A.~A.~C. and {Shenar}, T. and {Tramper}, F. and {Villase{\~n}or}, J.~I. and {Waisberg}, I.},
        title = "{Investigating 39 Galactic Wolf-Rayet stars with VLTI/GRAVITY: Uncovering a long-period binary desert}",
      journal = {\aap},
         year = 2024,
        month = dec,
       volume = {692},
          eid = {A109},
        pages = {A109},
          doi = {10.1051/0004-6361/202452352},
       adsurl = {https://ui.adsabs.harvard.edu/abs/2024A&A...692A.109D}
}

@ARTICLE{2009StLouis,
       author = {{St-Louis}, N. and {Chen{\'e}}, A. -N. and {Schnurr}, O. and
         {Nicol}, M. -H.},
        title = "{A Systematic Search for Corotating Interaction Regions in Apparently Single Galactic Wolf-Rayet Stars. I. Characterizing the Variability}",
      journal = {\apj},
         year = "2009",
        month = "Jun",
       volume = {698},
       number = {2},
        pages = {1951-1962},
          doi = {10.1088/0004-637X/698/2/1951},
archivePrefix = {arXiv},
       eprint = {0905.4460},
 primaryClass = {astro-ph.SR},
       adsurl = {https://ui.adsabs.harvard.edu/abs/2009ApJ...698.1951S}
}

@INPROCEEDINGS{1976Conti,
       author = {{Conti}, P.S.},
        title = "{}",
    booktitle = {Proc. 20th Colloq. Int. Astrophys.},
         year = "1976",
       editor = {},
       series = {Univ. Liege},
       volume = {},
        month = "",
        pages = {193},
       adsurl = {}
}

@ARTICLE{2015SmetteMolecfit,
       author = {{Smette}, A. and {Sana}, H. and {Noll}, S. and {Horst}, H. and
         {Kausch}, W. and {Kimeswenger}, S. and {Barden}, M. and {Szyszka}, C. and
         {Jones}, A.~M. and {Gallenne}, A. and {Vinther}, J. and
         {Ballester}, P. and {Taylor}, J.},
        title = "{Molecfit: A general tool for telluric absorption correction. I. Method and application to ESO instruments}",
      journal = {\aap},
         year = "2015",
        month = "Apr",
       volume = {576},
          eid = {A77},
        pages = {A77},
          doi = {10.1051/0004-6361/201423932},
archivePrefix = {arXiv},
       eprint = {1501.07239},
 primaryClass = {astro-ph.IM},
       adsurl = {https://ui.adsabs.harvard.edu/abs/2015A&A...576A..77S}
}

@ARTICLE{2020Sander,
       author = {{Sander}, Andreas A.~C. and {Vink}, J.~S. and {Hamann}, W. -R.},
        title = "{Driving classical Wolf-Rayet winds: a {\ensuremath{\Gamma}}- and Z-dependent mass-loss}",
      journal = {\mnras},
         year = 2020,
        month = jan,
       volume = {491},
       number = {3},
        pages = {4406-4425},
          doi = {10.1093/mnras/stz3064},
archivePrefix = {arXiv},
       eprint = {1910.12886},
 primaryClass = {astro-ph.SR},
       adsurl = {https://ui.adsabs.harvard.edu/abs/2020MNRAS.491.4406S}
}

@ARTICLE{2020Shenar,
       author = {{Shenar}, T. and {Gilkis}, A. and {Vink}, J.~S. and {Sana}, H. and {Sander}, A.~A.~C.},
        title = "{Why binary interaction does not necessarily dominate the formation of Wolf-Rayet stars at low metallicity}",
      journal = {\aap},
         year = "2020",
        month = "Feb",
       volume = {634},
          eid = {A79},
        pages = {A79},
          doi = {10.1051/0004-6361/201936948},
       adsurl = {https://ui.adsabs.harvard.edu/abs/2020A&A...634A..79S}
}

@ARTICLE{1967Paczynski,
       author = {{Paczy{\'n}ski}, B.},
        title = "{Evolution of Close Binaries. V. The Evolution of Massive Binaries and the Formation of the Wolf-Rayet Stars}",
      journal = {\actaa},
         year = 1967,
        month = jan,
       volume = {17},
        pages = {355},
       adsurl = {https://ui.adsabs.harvard.edu/abs/1967AcA....17..355P}
}

@ARTICLE{2019Woosley,
       author = {{Woosley}, S.~E.},
        title = "{The Evolution of Massive Helium Stars, Including Mass Loss}",
      journal = {\apj},
         year = 2019,
        month = jun,
       volume = {878},
       number = {1},
          eid = {49},
        pages = {49},
          doi = {10.3847/1538-4357/ab1b41},
archivePrefix = {arXiv},
       eprint = {1901.00215},
 primaryClass = {astro-ph.SR},
       adsurl = {https://ui.adsabs.harvard.edu/abs/2019ApJ...878...49W}
}

@ARTICLE{2013StLouis,
       author = {{St-Louis}, N.},
        title = "{Revealing the Asymmetry of the Wind of the Variable Wolf-Rayet Star WR1 (HD 4004) through Spectropolarization}",
      journal = {\apj},
         year = 2013,
        month = nov,
       volume = {777},
       number = {1},
          eid = {9},
        pages = {9},
          doi = {10.1088/0004-637X/777/1/9},
archivePrefix = {arXiv},
       eprint = {1308.5914},
 primaryClass = {astro-ph.SR},
       adsurl = {https://ui.adsabs.harvard.edu/abs/2013ApJ...777....9S}
}

@ARTICLE{2020Koenigsberger,
       author = {{Koenigsberger}, G. and {Schmutz}, W.},
        title = "{The nature of the companion in the Wolf-Rayet system EZ Canis Majoris}",
      journal = {\aap},
         year = 2020,
        month = jul,
       volume = {639},
          eid = {A18},
        pages = {A18},
          doi = {10.1051/0004-6361/201937305},
archivePrefix = {arXiv},
       eprint = {2005.06028},
 primaryClass = {astro-ph.SR},
       adsurl = {https://ui.adsabs.harvard.edu/abs/2020A&A...639A..18K}
}

@ARTICLE{1997Morel,
       author = {{Morel}, Thierry and {St-Louis}, Nicole and {Marchenko}, Sergey V.},
        title = "{Optical Spectroscopy of EZ Canis Majoris: Indication for Large-Scale Structures in a Wolf-Rayet Wind}",
      journal = {\apj},
         year = 1997,
        month = jun,
       volume = {482},
       number = {1},
        pages = {470-489},
          doi = {10.1086/304122},
       adsurl = {https://ui.adsabs.harvard.edu/abs/1997ApJ...482..470M}
}

@ARTICLE{2019SchmutzKoenigsbergerWR6,
       author = {{Schmutz}, W. and {Koenigsberger}, G.},
        title = "{Long uninterrupted photometric observations of the Wolf-Rayet star EZ CMa by the Toronto BRITE satellite reveal a very fast apsidal motion}",
      journal = {\aap},
         year = 2019,
        month = apr,
       volume = {624},
          eid = {L3},
        pages = {L3},
          doi = {10.1051/0004-6361/201935094},
archivePrefix = {arXiv},
       eprint = {1903.09501},
 primaryClass = {astro-ph.SR},
       adsurl = {https://ui.adsabs.harvard.edu/abs/2019A&A...624L...3S}
}

@ARTICLE{2004HamannGrafenerWN,
       author = {{Hamann}, W. -R. and {Gr{\"a}fener}, G.},
        title = "{Grids of model spectra for WN stars, ready for use}",
      journal = {\aap},
         year = 2004,
        month = nov,
       volume = {427},
        pages = {697-704},
          doi = {10.1051/0004-6361:20040506},
       adsurl = {https://ui.adsabs.harvard.edu/abs/2004A&A...427..697H}
}

@ARTICLE{2019Hamann,
       author = {{Hamann}, W. -R. and {Gr{\"a}fener}, G. and {Liermann}, A. and {Hainich}, R. and {Sander}, A.~A.~C. and {Shenar}, T. and {Ramachandran}, V. and {Todt}, H. and {Oskinova}, L.~M.},
        title = "{The Galactic WN stars revisited. Impact of Gaia distances on fundamental stellar parameters}",
      journal = {\aap},
         year = 2019,
        month = may,
       volume = {625},
          eid = {A57},
        pages = {A57},
          doi = {10.1051/0004-6361/201834850},
archivePrefix = {arXiv},
       eprint = {1904.04687},
 primaryClass = {astro-ph.SR},
       adsurl = {https://ui.adsabs.harvard.edu/abs/2019A&A...625A..57H}
}

@INPROCEEDINGS{2014Lapeyrere,
       author = {{Lapeyrere}, V. and {Kervella}, P. and {Lacour}, S. and {Azouaoui}, N. and {Garcia-Dabo}, C.~E. and {Perrin}, G. and {Eisenhauer}, F. and {Perraut}, K. and {Straubmeier}, C. and {Amorim}, A. and {Brandner}, W.},
        title = "{GRAVITY data reduction software}",
    booktitle = {Optical and Infrared Interferometry IV},
         year = 2014,
       editor = {{Rajagopal}, Jayadev K. and {Creech-Eakman}, Michelle J. and {Malbet}, Fabien},
       series = {Society of Photo-Optical Instrumentation Engineers (SPIE) Conference Series},
       volume = {9146},
        month = jul,
          eid = {91462D},
        pages = {91462D},
          doi = {10.1117/12.2056850},
       adsurl = {https://ui.adsabs.harvard.edu/abs/2014SPIE.9146E..2DL}
}

@ARTICLE{2023Lefever,
       author = {{Lefever}, R.~R. and {Sander}, A.~A.~C. and {Shenar}, T. and {Poniatowski}, L.~G. and {Dsilva}, K. and {Todt}, H.},
        title = "{Exploring the influence of different velocity fields on Wolf-Rayet star spectra}",
      journal = {\mnras},
         year = 2023,
        month = may,
       volume = {521},
       number = {1},
        pages = {1374-1392},
          doi = {10.1093/mnras/stad625},
archivePrefix = {arXiv},
       eprint = {2302.13964},
 primaryClass = {astro-ph.SR},
       adsurl = {https://ui.adsabs.harvard.edu/abs/2023MNRAS.521.1374L}
}

@ARTICLE{2013Pecaut,
       author = {{Pecaut}, Mark J. and {Mamajek}, Eric E.},
        title = "{Intrinsic Colors, Temperatures, and Bolometric Corrections of Pre-main-sequence Stars}",
      journal = {\apjs},
         year = 2013,
        month = sep,
       volume = {208},
       number = {1},
          eid = {9},
        pages = {9},
          doi = {10.1088/0067-0049/208/1/9},
archivePrefix = {arXiv},
       eprint = {1307.2657},
 primaryClass = {astro-ph.SR},
       adsurl = {https://ui.adsabs.harvard.edu/abs/2013ApJS..208....9P}
}

@ARTICLE{2011Chene,
       author = {{Chen{\'e}}, A. -N. and {Moffat}, A.~F.~J. and {Cameron}, C. and {Fahed}, R. and {Gamen}, R.~C. and {Lef{\`e}vre}, L. and {Rowe}, J.~F. and {St-louis}, N. and {Muntean}, V. and {De La Chevroti{\`e}re}, A. and {Guenther}, D.~B. and {Kuschnig}, R. and {Matthews}, J.~M. and {Rucinski}, S.~M. and {Sasselov}, D. and {Weiss}, W.~W.},
        title = "{WR 110: A Single Wolf-Rayet Star with Corotating Interaction Regions in its Wind?}",
      journal = {\apj},
         year = 2011,
        month = jul,
       volume = {735},
       number = {1},
          eid = {34},
        pages = {34},
          doi = {10.1088/0004-637X/735/1/34},
archivePrefix = {arXiv},
       eprint = {1105.0919},
 primaryClass = {astro-ph.SR},
       adsurl = {https://ui.adsabs.harvard.edu/abs/2011ApJ...735...34C}
}

@ARTICLE{2014Smith,
       author = {{Smith}, Nathan},
        title = "{Mass Loss: Its Effect on the Evolution and Fate of High-Mass Stars}",
      journal = {\araa},
         year = 2014,
        month = aug,
       volume = {52},
        pages = {487-528},
          doi = {10.1146/annurev-astro-081913-040025},
archivePrefix = {arXiv},
       eprint = {1402.1237},
 primaryClass = {astro-ph.SR},
       adsurl = {https://ui.adsabs.harvard.edu/abs/2014ARA&A..52..487S}
}

@ARTICLE{2015Gallenne,
       author = {{Gallenne}, A. and {M{\'e}rand}, A. and {Kervella}, P. and {Monnier}, J.~D. and {Schaefer}, G.~H. and {Baron}, F. and {Breitfelder}, J. and {Le Bouquin}, J.~B. and {Roettenbacher}, R.~M. and {Gieren}, W. and {Pietrzy{\'n}ski}, G. and {McAlister}, H. and {ten Brummelaar}, T. and {Sturmann}, J. and {Sturmann}, L. and {Turner}, N. and {Ridgway}, S. and {Kraus}, S.},
        title = "{Robust high-contrast companion detection from interferometric observations. The CANDID algorithm and an application to six binary Cepheids}",
      journal = {\aap},
         year = 2015,
        month = jul,
       volume = {579},
          eid = {A68},
        pages = {A68},
          doi = {10.1051/0004-6361/201525917},
archivePrefix = {arXiv},
       eprint = {1505.02715},
 primaryClass = {astro-ph.SR},
       adsurl = {https://ui.adsabs.harvard.edu/abs/2015A&A...579A..68G}
}

@ARTICLE{2023Sander,
       author = {{Sander}, A.~A.~C. and {Lefever}, R.~R. and {Poniatowski}, L.~G. and {Ramachandran}, V. and {Sabhahit}, G.~N. and {Vink}, J.~S.},
        title = "{The temperature dependency of Wolf-Rayet-type mass loss. An exploratory study for winds launched by the hot iron bump}",
      journal = {\aap},
         year = 2023,
        month = feb,
       volume = {670},
          eid = {A83},
        pages = {A83},
          doi = {10.1051/0004-6361/202245110},
archivePrefix = {arXiv},
       eprint = {2301.01785},
 primaryClass = {astro-ph.SR},
       adsurl = {https://ui.adsabs.harvard.edu/abs/2023A&A...670A..83S}
}

@ARTICLE{1987Hillier,
       author = {{Hillier}, D.~J.},
        title = "{An Empirical Model for the Wolf-Rayet Star HD 50896}",
      journal = {\apjs},
         year = 1987,
        month = apr,
       volume = {63},
        pages = {965},
          doi = {10.1086/191188},
       adsurl = {https://ui.adsabs.harvard.edu/abs/1987ApJS...63..965H}
}

@INPROCEEDINGS{2005Kennicutt,
       author = {{Kennicutt}, Robert C.},
        title = "{The role of massive stars in astrophysics}",
    booktitle = {Massive Star Birth: A Crossroads of Astrophysics},
         year = 2005,
       editor = {{Cesaroni}, R. and {Felli}, M. and {Churchwell}, E. and {Walmsley}, M.},
       series = {IAU Symposium},
       volume = {227},
        month = jan,
        pages = {3-11},
          doi = {10.1017/S1743921305004308},
       adsurl = {https://ui.adsabs.harvard.edu/abs/2005IAUS..227....3K}
}

@ARTICLE{1980Firmani,
       author = {{Firmani}, C. and {Koenigsberger}, G. and {Bisiacchi}, G.~F. and {Moffat}, A.~F.~J. and {Isserstedt}, J.},
        title = "{The binary nature of the single-line Wolf-Rayet star EZ CMa=HD 50896.}",
      journal = {\apj},
         year = 1980,
        month = jul,
       volume = {239},
        pages = {607-621},
          doi = {10.1086/158148},
       adsurl = {https://ui.adsabs.harvard.edu/abs/1980ApJ...239..607F}
}

@ARTICLE{1989Howarth,
       author = {{Howarth}, Ian D. and {Prinja}, Raman K.},
        title = "{The Stellar Winds of 203 Galactic O Stars: A Quantitative Ultraviolet Survey}",
      journal = {\apjs},
         year = 1989,
        month = mar,
       volume = {69},
        pages = {527},
          doi = {10.1086/191321},
       adsurl = {https://ui.adsabs.harvard.edu/abs/1989ApJS...69..527H}
}

@ARTICLE{1996Cranmer,
       author = {{Cranmer}, Steven R. and {Owocki}, Stanley P.},
        title = "{Hydrodynamical Simulations of Corotating Interaction Regions and Discrete Absorption Components in Rotating O-Star Winds}",
      journal = {\apj},
         year = 1996,
        month = may,
       volume = {462},
        pages = {469},
          doi = {10.1086/177166},
archivePrefix = {arXiv},
       eprint = {astro-ph/9508004},
 primaryClass = {astro-ph},
       adsurl = {https://ui.adsabs.harvard.edu/abs/1996ApJ...462..469C}
}

@ARTICLE{2023Ignace,
       author = {{Ignace}, R. and {Bjorkman}, J.~E. and {Chen{\'e}}, A.-N. and {Erba}, C. and {Fabiani}, L. and {Moffat}, A.~F.~J. and {Sincennes}, R. and {St-Louis}, N.},
        title = "{Modelling variable linear polarization produced by Co-rotating Interaction Regions (CIRs) across optical recombination lines of Wolf-Rayet stars}",
      journal = {\mnras},
         year = 2023,
        month = nov,
       volume = {526},
       number = {1},
        pages = {1298-1307},
          doi = {10.1093/mnras/stad2878},
archivePrefix = {arXiv},
       eprint = {2309.10599},
 primaryClass = {astro-ph.SR},
       adsurl = {https://ui.adsabs.harvard.edu/abs/2023MNRAS.526.1298I}
}

@ARTICLE{2016Aldoretta,
       author = {{Aldoretta}, E.~J. and {St-Louis}, N. and {Richardson}, N.~D. and {Moffat}, A.~F.~J. and {Eversberg}, T. and {Hill}, G.~M. and {Shenar}, T. and {Artigau}, {\'E}. and {Gauza}, B. and {Knapen}, J.~H. and {Kub{\'a}t}, J. and {Kub{\'a}tov{\'a}}, B. and {Maltais-Tariant}, R. and {Mu{\~n}oz}, M. and {Pablo}, H. and {Ramiaramanantsoa}, T. and {Richard-Laferri{\`e}re}, A. and {Sablowski}, D.~P. and {Sim{\'o}n-D{\'\i}az}, S. and {St-Jean}, L. and {Bolduan}, F. and {Dias}, F.~M. and {Dubreuil}, P. and {Fuchs}, D. and {Garrel}, T. and {Grutzeck}, G. and {Hunger}, T. and {K{\"u}sters}, D. and {Langenbrink}, M. and {Leadbeater}, R. and {Li}, D. and {Lopez}, A. and {Mauclaire}, B. and {Moldenhawer}, T. and {Potter}, M. and {dos Santos}, E.~M. and {Schanne}, L. and {Schmidt}, J. and {Sieske}, H. and {Strachan}, J. and {Stinner}, E. and {Stinner}, P. and {Stober}, B. and {Strandbaek}, K. and {Syder}, T. and {Verilhac}, D. and {Waldschl{\"a}ger}, U. and {Weiss}, D. and {Wendt}, A.},
        title = "{An extensive spectroscopic time series of three Wolf-Rayet stars - I. The lifetime of large-scale structures in the wind of WR 134}",
      journal = {\mnras},
         year = 2016,
        month = aug,
       volume = {460},
       number = {3},
        pages = {3407-3417},
          doi = {10.1093/mnras/stw1188},
archivePrefix = {arXiv},
       eprint = {1605.04868},
 primaryClass = {astro-ph.SR},
       adsurl = {https://ui.adsabs.harvard.edu/abs/2016MNRAS.460.3407A}
}

@ARTICLE{2010Chene,
       author = {{Chen{\'e}}, A.-N. and {St-Louis}, N.},
        title = "{Large-scale Periodic Variability of the Wind of the Wolf-Rayet Star WR 1 (HD 4004)}",
      journal = {\apj},
         year = 2010,
        month = jun,
       volume = {716},
       number = {2},
        pages = {929-941},
          doi = {10.1088/0004-637X/716/2/929},
archivePrefix = {arXiv},
       eprint = {1004.4893},
 primaryClass = {astro-ph.SR},
       adsurl = {https://ui.adsabs.harvard.edu/abs/2010ApJ...716..929C}
}

@ARTICLE{1999bMorel,
       author = {{Morel}, Thierry and {Marchenko}, Sergey V. and {Eenens}, Philippe R.~J. and {Moffat}, Anthony F.~J. and {Koenigsberger}, Gloria and {Antokhin}, Igor I. and {Eversberg}, Thomas and {Tovmassian}, Gaghik H. and {Hill}, Grant M. and {Cardona}, Octavio and {St-Louis}, Nicole},
        title = "{A 2.3 Day Periodic Variability in the Apparently Single Wolf-Rayet Star WR 134: Collapsed Companion or Rotational Modulation?}",
      journal = {\apj},
         year = 1999,
        month = jun,
       volume = {518},
       number = {1},
        pages = {428-441},
          doi = {10.1086/307250},
archivePrefix = {arXiv},
       eprint = {astro-ph/9901269},
 primaryClass = {astro-ph},
       adsurl = {https://ui.adsabs.harvard.edu/abs/1999ApJ...518..428M}
}

@ARTICLE{RNAAS2025,
       author = {{Steenken}, Nicolaus and {Ignace}, Richard and {St-Louis}, Nicole and {Lenoir-Craig}, Guillaume},
        title = "{Optical Polarimetry of WR 1 Consistent with a CIR and 17 days Rotation}",
      journal = {Research Notes of the American Astronomical Society},
         year = 2025,
        month = feb,
       volume = {9},
       number = {2},
          eid = {36},
        pages = {36},
          doi = {10.3847/2515-5172/adb4ea},
       adsurl = {https://ui.adsabs.harvard.edu/abs/2025RNAAS...9...36S}
}

@ARTICLE{2002Dessart,
       author = {{Dessart}, L. and {Chesneau}, O.},
        title = "{Interferometric and spectroscopic monitoring of emission lines. Detection of CIRs in hot star winds}",
      journal = {\aap},
         year = 2002,
        month = nov,
       volume = {395},
        pages = {209-221},
          doi = {10.1051/0004-6361:20021110},
       adsurl = {https://ui.adsabs.harvard.edu/abs/2002A&A...395..209D}
}

@ARTICLE{1984Mullan,
       author = {{Mullan}, D.~J.},
        title = "{Corotating interaction regions in stellar winds}",
      journal = {\apj},
         year = 1984,
        month = aug,
       volume = {283},
        pages = {303-312},
          doi = {10.1086/162307},
       adsurl = {https://ui.adsabs.harvard.edu/abs/1984ApJ...283..303M}
}

@ARTICLE{2020Chene,
       author = {{Chen{\'e}}, Andr{\'e}-Nicolas and {St-Louis}, Nicole and {Moffat}, Anthony F.~J. and {Gayley}, Kenneth G.},
        title = "{Clumping in the Winds of Wolf-Rayet Stars}",
      journal = {\apj},
         year = 2020,
        month = nov,
       volume = {903},
       number = {2},
          eid = {113},
        pages = {113},
          doi = {10.3847/1538-4357/abba24},
archivePrefix = {arXiv},
       eprint = {2009.10341},
 primaryClass = {astro-ph.SR},
       adsurl = {https://ui.adsabs.harvard.edu/abs/2020ApJ...903..113C}
}

@ARTICLE{2002Skinner,
       author = {{Skinner}, Stephen L. and {Zhekov}, Svetozar A. and {G{\"u}del}, Manuel and {Schmutz}, Werner},
        title = "{XMM-Newton and Very Large Array Observations of the Variable Wolf-Rayet Star EZ Canis Majoris: Evidence for a Close Companion?}",
      journal = {\apj},
         year = 2002,
        month = nov,
       volume = {579},
       number = {2},
        pages = {764-773},
          doi = {10.1086/342841},
archivePrefix = {arXiv},
       eprint = {astro-ph/0207171},
 primaryClass = {astro-ph},
       adsurl = {https://ui.adsabs.harvard.edu/abs/2002ApJ...579..764S}
}

@ARTICLE{2026Fabiani,
       author = {{Fabiani}, Luca and {St-Louis}, Nicole and {Moffat}, Anthony F. J. and {Chen\'e}, Andr\'e-Nicolas and {Ignace}, {Richard}},
        title = "{Variable linear polarisation across optical emission lines of the Wolf-Rayet star EZ Canis Majoris and its relation to Corotating Interaction Regions}",
      journal = {\apj},
         year = 2026,
        month = nxx,
       volume = {in},
       number = {press},
        pages = {press},
          doi = {xx},
       adsurl = {xx}
}

@ARTICLE{2024Freudling,
       author = {{Freudling}, W. and {Zampieri}, S. and {Coccato}, L. and {Podgorski}, S. and {Romaniello}, M. and {Modigliani}, A. and {Pritchard}, J.},
        title = "{Adaptive data reduction workflows for astronomy: The ESO Data Processing System (EDPS)}",
      journal = {\aap},
         year = 2024,
        month = jan,
       volume = {681},
          eid = {A93},
        pages = {A93},
          doi = {10.1051/0004-6361/202347651},
archivePrefix = {arXiv},
       eprint = {2311.03822},
 primaryClass = {astro-ph.IM},
       adsurl = {https://ui.adsabs.harvard.edu/abs/2024A&A...681A..93F}
}

@ARTICLE{2023Crowther,
       author = {{Crowther}, Paul A. and {Rate}, G. and {Bestenlehner}, Joachim M.},
        title = "{Line luminosities of Galactic and Magellanic Cloud Wolf-Rayet stars}",
      journal = {\mnras},
         year = 2023,
        month = may,
       volume = {521},
       number = {1},
        pages = {585-612},
          doi = {10.1093/mnras/stad418},
archivePrefix = {arXiv},
       eprint = {2301.11297},
 primaryClass = {astro-ph.SR},
       adsurl = {https://ui.adsabs.harvard.edu/abs/2023MNRAS.521..585C}
}

@ARTICLE{1983ApJ...271..221H,
       author = {{Hillier}, D.~J. and {Jones}, T.~J. and {Hyland}, A.~R.},
        title = "{Infrared spectra of WN stars. I. HD 50896.}",
      journal = {\apj},
         year = 1983,
        month = aug,
       volume = {271},
        pages = {221-236},
          doi = {10.1086/161189},
       adsurl = {https://ui.adsabs.harvard.edu/abs/1983ApJ...271..221H}
}

@ARTICLE{1997ApJ...486..420F,
       author = {{Figer}, Donald F. and {McLean}, Ian S. and {Najarro}, Francisco},
        title = "{A K-Band Spectral Atlas of Wolf-Rayet Stars}",
      journal = {\apj},
         year = 1997,
        month = sep,
       volume = {486},
       number = {1},
        pages = {420-434},
          doi = {10.1086/304488},
       adsurl = {https://ui.adsabs.harvard.edu/abs/1997ApJ...486..420F}
}

@ARTICLE{2011Bonneau,
       author = {{Bonneau}, D. and {Delfosse}, X. and {Mourard}, D. and {Lafrasse}, S. and {Mella}, G. and {Cetre}, S. and {Clausse}, J.-M. and {Zins}, G.},
        title = "{SearchCal: a Virtual Observatory tool for searching calibrators in optical long-baseline interferometry. II. The faint-object case}",
      journal = {\aap},
         year = 2011,
        month = nov,
       volume = {535},
          eid = {A53},
        pages = {A53},
          doi = {10.1051/0004-6361/201015124},
       adsurl = {https://ui.adsabs.harvard.edu/abs/2011A&A...535A..53B}
}

@ARTICLE{2020Speagle,
       author = {{Speagle}, Joshua S.},
        title = "{DYNESTY: a dynamic nested sampling package for estimating Bayesian posteriors and evidences}",
      journal = {\mnras},
         year = 2020,
        month = apr,
       volume = {493},
       number = {3},
        pages = {3132-3158},
          doi = {10.1093/mnras/staa278},
archivePrefix = {arXiv},
       eprint = {1904.02180},
 primaryClass = {astro-ph.IM},
       adsurl = {https://ui.adsabs.harvard.edu/abs/2020MNRAS.493.3132S}
}

@INPROCEEDINGS{2004Skilling,
       author = {{Skilling}, John},
        title = "{Nested Sampling}",
    booktitle = {Bayesian Inference and Maximum Entropy Methods in Science and Engineering: 24th International Workshop on Bayesian Inference and Maximum Entropy Methods in Science and Engineering},
         year = 2004,
       editor = {{Fischer}, Rainer and {Preuss}, Roland and {Toussaint}, Udo Von},
       series = {American Institute of Physics Conference Series},
       volume = {735},
        month = nov,
    publisher = {AIP},
        pages = {395-405},
          doi = {10.1063/1.1835238},
       adsurl = {https://ui.adsabs.harvard.edu/abs/2004AIPC..735..395S}
}

@article{1963Marquardt,
 ISSN = {03684245},
 URL = {http://www.jstor.org/stable/2098941},
 author = {Donald W. Marquardt},
 journal = {Journal of the Society for Industrial and Applied Mathematics},
 number = {2},
 pages = {431--441},
 publisher = {Society for Industrial and Applied Mathematics},
 title = {An Algorithm for Least-Squares Estimation of Nonlinear Parameters},
 urldate = {2026-09-09},
 volume = {11},
 year = {1963}
}

@article{1979Efron,
  author = {Efron, B.},
  doi = {10.1214/aos/1176344552},
  journal = {The Annals of Statistics},
  month = jan,
  number = 1,
  pages = {1--26},
  title = {Bootstrap Methods: Another Look at the Jackknife},
  url = {https://doi.org/10.1214/aos/1176344552},
  volume = 7,
  year = 1979
}

@ARTICLE{2017Sander,
       author = {{Sander}, A.~A.~C. and {Hamann}, W.-R. and {Todt}, H. and {Hainich}, R. and {Shenar}, T.},
        title = "{Coupling hydrodynamics with comoving frame radiative transfer. I. A unified approach for OB and WR stars}",
      journal = {\aap},
         year = 2017,
        month = jul,
       volume = {603},
          eid = {A86},
        pages = {A86},
          doi = {10.1051/0004-6361/201730642},
archivePrefix = {arXiv},
       eprint = {1704.08698},
 primaryClass = {astro-ph.SR},
       adsurl = {https://ui.adsabs.harvard.edu/abs/2017A&A...603A..86S}
}

@ARTICLE{2026Lefever,
       author = {{Lefever}, R.~R. and {Sander}, A.~A.~C. and {Bernini-Peron}, M. and {Gonz{\'a}lez-Tor{\`a}}, G. and {Hamann}, W.-R. and {Josiek}, J. and {Ramachandran}, V. and {Sch{\"o}sser}, E.~C. and {Todt}, H.},
        title = "{Dynamically consistent analysis of Galactic WN4b stars: Revised parameters and insights on Wolf─Rayet mass-loss treatments}",
      journal = {\aap},
         year = 2026,
        month = feb,
       volume = {707},
          eid = {A8},
        pages = {A8},
          doi = {10.1051/0004-6361/202558211},
archivePrefix = {arXiv},
       eprint = {2601.02498},
 primaryClass = {astro-ph.SR},
       adsurl = {https://ui.adsabs.harvard.edu/abs/2026A&A...707A...8L}
}

@ARTICLE{1988Hillier,
       author = {{Hillier}, D.~J.},
        title = "{The Formation of Nitrogen and Carbon Emission Lines in HD 50896 (WN5)}",
      journal = {\apj},
         year = 1988,
        month = apr,
       volume = {327},
        pages = {822},
          doi = {10.1086/166240},
       adsurl = {https://ui.adsabs.harvard.edu/abs/1988ApJ...327..822H}
}

@ARTICLE{2018Toala,
       author = {{Toal{\'a}}, J.~A. and {Oskinova}, L.~M. and {Hamann}, W.-R. and {Ignace}, R. and {Sander}, A.~A.~C. and {Shenar}, T. and {Todt}, H. and {Chu}, Y.-H. and {Guerrero}, M.~A. and {Hainich}, R. and {Torrej{\'o}n}, J.~M.},
        title = "{On the Apparent Absence of Wolf-Rayet+Neutron Star Systems: The Curious Case of WR124}",
      journal = {\apjl},
         year = 2018,
        month = dec,
       volume = {869},
       number = {1},
          eid = {L11},
        pages = {L11},
          doi = {10.3847/2041-8213/aaf39d},
archivePrefix = {arXiv},
       eprint = {1811.10691},
 primaryClass = {astro-ph.HE},
       adsurl = {https://ui.adsabs.harvard.edu/abs/2018ApJ...869L..11T}
}

@ARTICLE{2000Contreras,
       author = {{Contreras}, M.~E. and {Rodr{\'\i}guez}, L.~F. and {Arnal}, E.~M.},
        title = "{New VLA Observations of WR 6 (= HR 50896): A Search for an Anisotropic Wind}",
      journal = {\rmxaa},
         year = 2000,
        month = oct,
       volume = {36},
        pages = {135-139},
       adsurl = {https://ui.adsabs.harvard.edu/abs/2000RMxAA..36..135C}
}

@ARTICLE{2024Barclay,
       author = {{Barclay}, Krister D.~G. and {Rosu}, Sophie and {Richardson}, Noel D. and {Chen{\'e}}, Andr{\'e}-Nicolas and {St-Louis}, Nicole and {Ignace}, Richard and {Moffat}, Anthony F.~J.},
        title = "{Using CHIRON spectroscopy to test the hypothesis of a precessing orbit for the WN4 star EZ CMa}",
      journal = {\mnras},
         year = 2024,
        month = jan,
       volume = {527},
       number = {2},
        pages = {2198-2208},
          doi = {10.1093/mnras/stad3289},
archivePrefix = {arXiv},
       eprint = {2310.15986},
 primaryClass = {astro-ph.SR},
       adsurl = {https://ui.adsabs.harvard.edu/abs/2024MNRAS.527.2198B}
}

@INPROCEEDINGS{2008Moffat,
       author = {{Moffat}, Anthony F.~J.},
        title = "{Observational overview of clumping in hot stellar winds}",
    booktitle = {Clumping in Hot-Star Winds},
         year = 2008,
       editor = {{Hamann}, Wolf-Rainer and {Feldmeier}, Achim and {Oskinova}, Lidia M.},
        month = apr,
        pages = {17},
       adsurl = {https://ui.adsabs.harvard.edu/abs/2008cihw.conf...17M}
}

\begin{appendix}

\counterwithin{figure}{section}

\section{Additional figures}
\label{sec:app}

Below we show the GRAVITY data and best-fit models for all epochs in a format similar to Figure\,\ref{fig:zoom}.

\begin{wrapfigure}{r}{0.999\textwidth} 
\label{fig:1zoom}
\centering
\includegraphics[trim={0cm 0cm 0cm 0cm}, clip,width=0.999\textwidth]{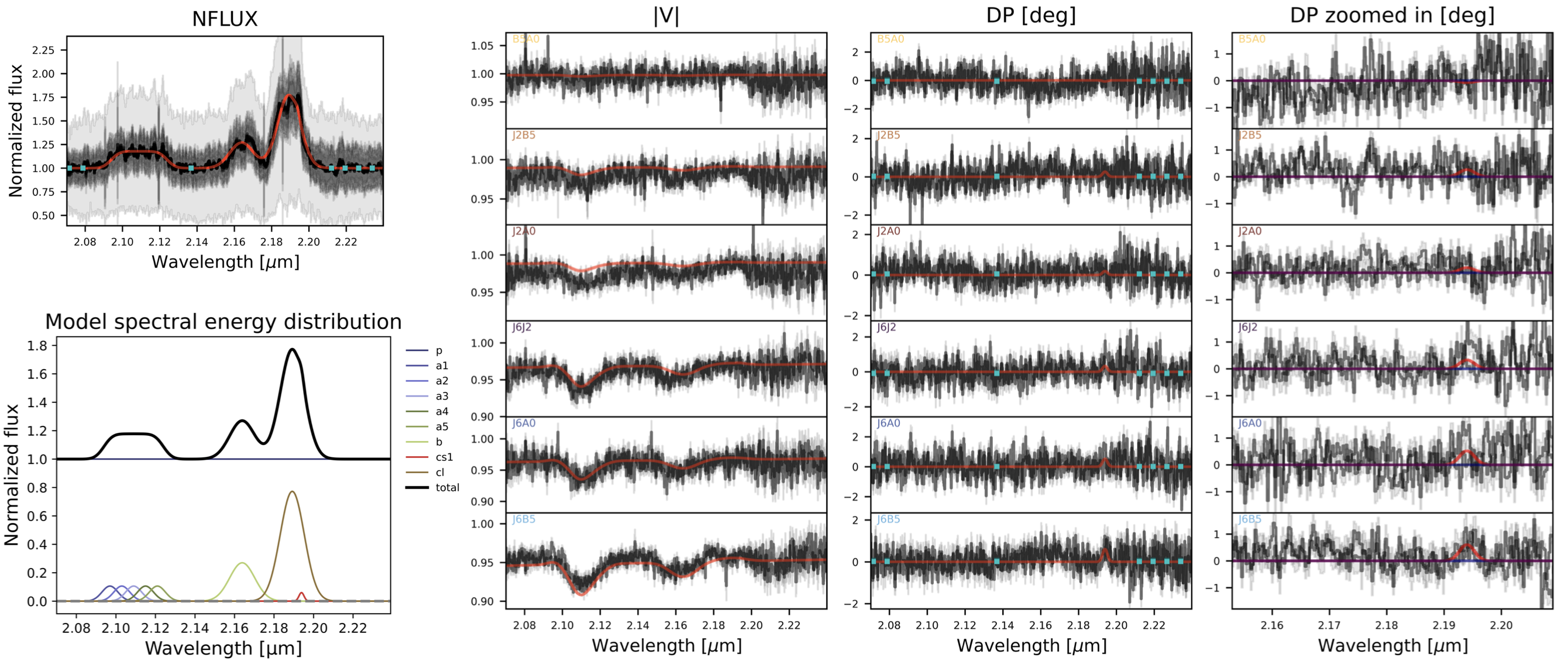}
\caption{Same plot as Figure\,\ref{fig:zoom} for Epoch 1.} 
\end{wrapfigure}

\begin{wrapfigure}{r}{0.999\textwidth} 
\label{fig:2zoom}
\centering
\includegraphics[trim={0cm 0cm 0cm 0cm}, clip,width=0.999\textwidth]{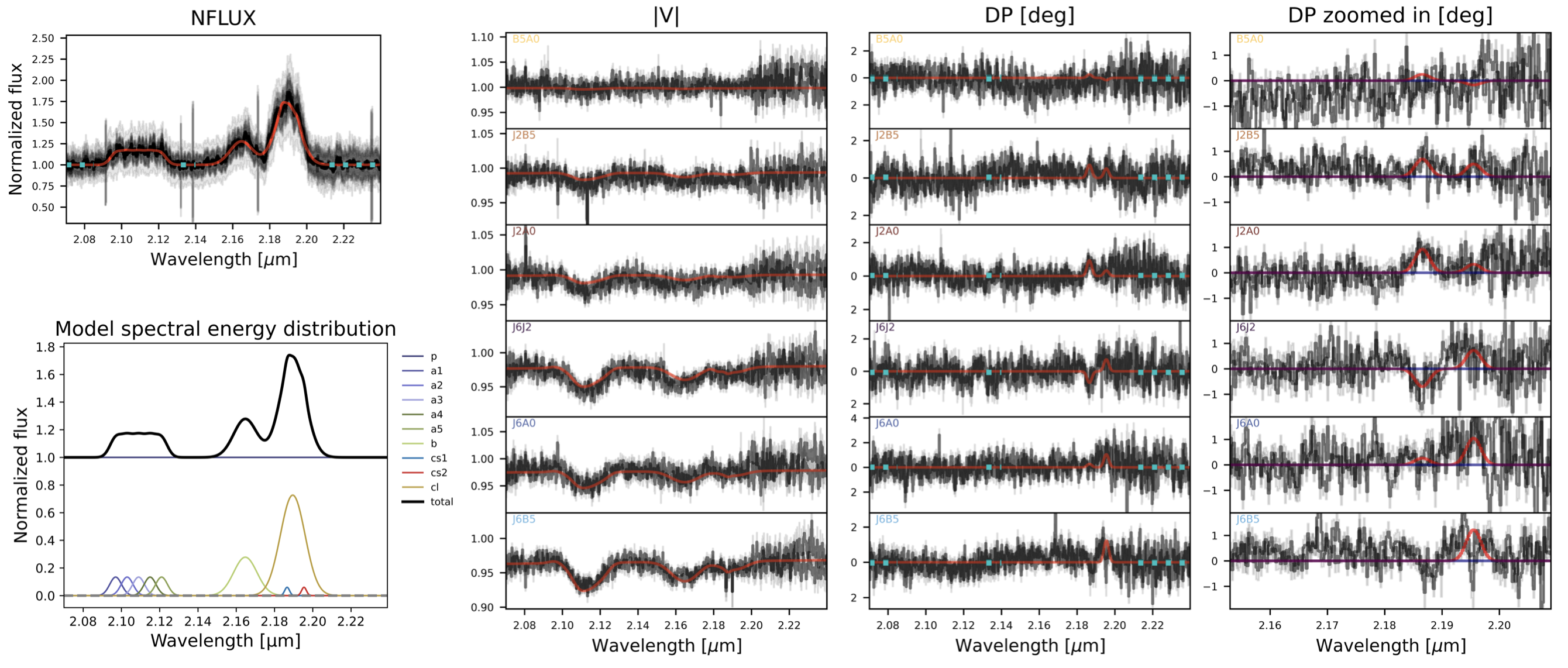}
\caption{Same plot as Figure\,\ref{fig:zoom} for Epoch 2.} 
\end{wrapfigure}

\clearpage

\begin{wrapfigure}{r}{0.999\textwidth} 
\label{fig:4zoom}
\centering
\includegraphics[trim={0cm 0cm 0cm 0cm}, clip,width=0.999\textwidth]{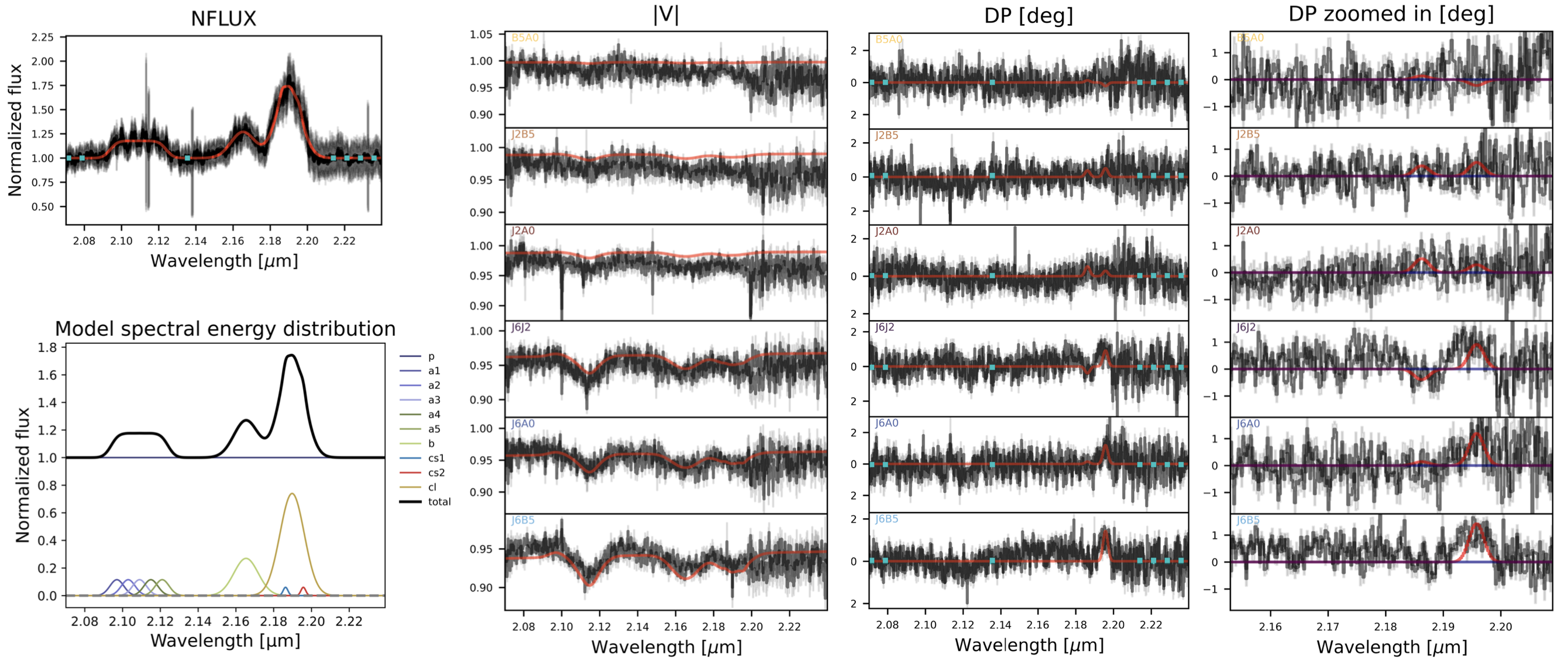}
\caption{Same plot as Figure\,\ref{fig:zoom} for Epoch 4.} 
\end{wrapfigure}

\begin{wrapfigure}{r}{0.999\textwidth} 
\label{fig:5zoom}
\centering
\includegraphics[trim={0cm 0cm 0cm 0cm}, clip,width=0.999\textwidth]{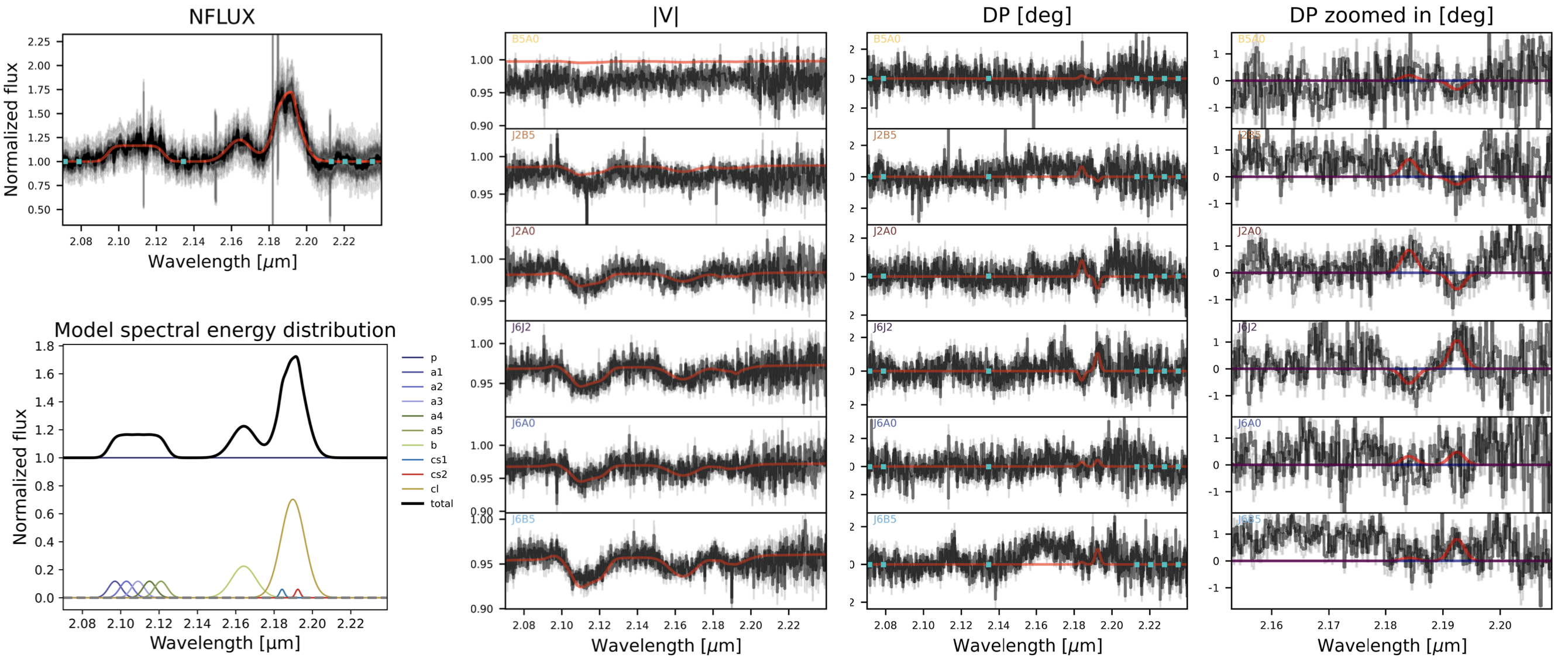}
\caption{Same plot as Figure\,\ref{fig:zoom} for Epoch 5.} 
\end{wrapfigure}

\clearpage

\begin{wrapfigure}{r}{0.999\textwidth} 
\label{fig:6zoom}
\centering
\includegraphics[trim={0cm 0cm 0cm 0cm}, clip,width=0.999\textwidth]{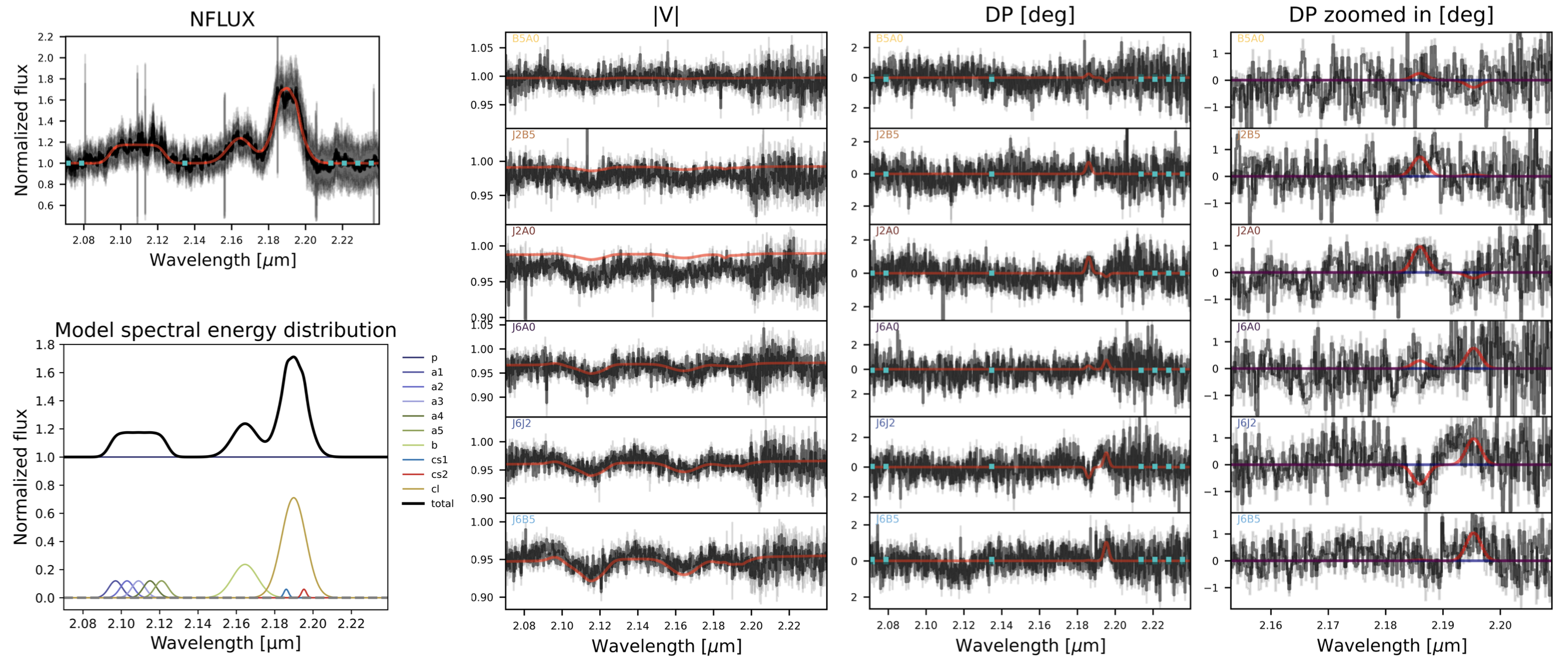}
\caption{Same plot as Figure\,\ref{fig:zoom} for Epoch 6.} 
\end{wrapfigure}

\end{appendix}

\end{document}